\documentclass[aps,pre,preprint,groupedaddress,longbibliography]{revtex4-1}
\usepackage[dvips]{graphicx}
\usepackage[utf8]{inputenc}
\usepackage{psfrag}
\usepackage{amsthm}
\usepackage{amsmath}
\usepackage{amssymb}

\newtheorem{Theorem}{Theorem}
\newtheorem{Definition}{Definition}
\newtheorem{Lemma}{Lemma}

\newcommand{\beqn}{\begin{eqnarray*}}
\newcommand{\eeqn}{\end{eqnarray*}}

\newcommand{\beq}{\begin{eqnarray}}
\newcommand{\eeq}{\end{eqnarray}}

\newcommand{\abs}[1]{\left\vert #1 \right\vert}
\newcommand{\norm}[1]{\left\Vert #1 \right\Vert}

\newcommand{\R}{{\mathbb R}}  

\begin{document}

\title{How to de-synchronize quorum-sensing networks}

\author{Giovanni Russo}

\affiliation{Giovanni Russo is a Research Staff Member in Optimization, Control and Decision Science, 
IBM Research Ireland, {\tt\small grusso@ie.ibm.com}}

\date{\today}

\begin{abstract}
In this paper we investigate how so-called quorum-sensing networks can be de-synchronized. Such networks, which arise in many important application fields such as systems biology, are characterized by the fact that direct communication between network nodes is superimposed to  communication with a shared, environmental, variable. In particular, we provide a new sufficient condition ensuring that the trajectories of these quorum-sensing networks diverge from their synchronous evolution. Then, we apply our result to study two applications.\newline 
\vspace{0.5cm}
\noindent {\bf Preprint published in Physical Review E, 95, 042312, 2017.}
\end{abstract}


\maketitle

\section{INTRODUCTION}

The problem of studying the emerging behaviors in complex networks has attracted the attention of many scientists coming from different fields. A key motivation for this is that the study of these emerging dynamics is important for a number of applications, including social networks, \cite{Ini_09}, \cite{Li_06} and biology \cite{Gon_Ber_Wal_Kra_Her_05}, \cite{Ana_Mon_Bar_Buz_Koc_10}, \cite{Gol_09}.

Over the past few years, a large body of literature has been devoted to unveil the  mechanisms that are responsible of coordinated behaviors. Of particular interest among the physics community has been the study of a particular form of coordination: synchronization, see e.g. \cite{Rus_diB_09b}, \cite{Pec_Car_90}, \cite{Zha_15}, \cite{Gin_10}, \cite{Cho_10}. In such papers (and related references) several conditions have been devised ensuring that a network synchronizes. 

The common underlying assumption in many works on network synchronization is that nodes directly communicate with each other via some form of diffusive coupling. In many applications arising in networks from both nature and technology, however, this form of communication is often superimposed to a communication via a shared (environmental) variable. Bacteria, for instance, produce, release and sense signaling molecules. Such molecules can diffuse in the environment and are used by bacteria for population coordination. This mechanism is known as {\it quorum sensing}, \cite{Ng_Bas_09}. In a neuronal context, a  mechanism, where the coupling between individual network nodes (e.g. oscillators) is not direct but is
rather  implemented through a common medium, involves local field potentials \cite{Ana_Mon_Bar_Buz_Koc_10}, \cite{Per_Pez_Sah_Mit_And_02}. 

From a system dynamics viewpoint,  quorum-sensing networks have been recently studied in \cite{Rus_Slo_10}, where it has been shown that the shared environmental variable plays a key role for network synchronization by implementing a sort of distributed filter sensed as input by all network nodes.  We now address the different question of how these quorum-sensing networks can be de-synchronized. This is a relevant question in many application fields. For example, the loss of a coordinated behavior is sometimes synonymous of a  poor network design as it might cause amplification of disturbances and noise (see e.g. \cite{Bar_05}). In some other contexts, instead, de-synchronization is desirable. For instance,  it is believed that pathological synchronization among bursting neurons in the basal ganglia-cortical loop might be linked to the tremors seen in patients with Parkinson's disease, \cite{Wil_Moe_14}, \cite{Wil_Moh_14a}, \cite{Ahn_15}. 

\subsection*{Related Work}

In this Section, we now revise some works on network de-synchronization and quorum-sensing networks relevant for this paper. We also outline the main contributions of this paper in the context of the related Literature.

\noindent {\bf Quorum sensing.} Literature devoted to the study of the emerging behaviors in quorum sensing networks  (e.g., \cite{Gar_Elo_Str_04}, \cite{Tab_Slo_Pha_09}, \cite{Sak_13}, \cite{Kat_08}) is sparse when compared to that on diffusive topologies. Moreover, in some cases, results are obtained by neglecting the dynamics of the quorum/environmental variables, as well as the global effects of nonlinearities. This sparsity of results appears to be surprising as quorum-sensing mechanisms, besides their pervasiveness in natural systems, could also be used to somehow optimize the topology of technological networks. For example, the use of a shared variable significantly reduces the number of links required to achieve a given level of connectivity~\cite{Tab_Slo_Pha_09}.

\noindent {\bf De-synchronization.} A key technique to study network de-synchronization is the Master Stability Function (MSF) \cite{Pec_Car_98}, which provides a condition for de-synchronization based on the calculation of the  maximum  Floquet or  Lyapunov  exponents for  the  generic variational equation obtained from network dynamics (see also \cite{Hu_Yan_Liu_98}, \cite{Hua_Che_Lai_Pec_09}, \cite{Pec_00}, \cite{Fin_00}, \cite{Sor_12} and references therein). Recently the MSF approach has been also extended to the case of a global variable coupling the oscillators  and to the case of global coupling between nodes, see \cite{Yan_Lin_Wan_Hua_15}, \cite{Zan_Mik_98} and references therein. Finally, an approach to control de-synchronization has been presented in \cite{Wil_Moe_14}. In such a paper, the authors recast de-synchronization as an optimization problem. Other de-synchronization control methods include e.g. double-pulse stimulation, \cite{Tas_01}, nonlinear time-delayed feedback \cite{Kis_Rus_Kor_Hud_07}, phase resetting \cite{Dan_Hes_Moe_09}, \cite{Nab_Mir_Gib_Moe_13}. Also, in \cite{Dan_Nab_Moe_10}, an energy-optimal stimulus was used to control neural spike timing, while in \cite{Tal_Car_Kha_11}, a stimulation-based approach  has been developed to control synchrony in neural networks. Notable works on de-synchronization has also been carried in e.g. \cite{He_14}, \cite{Hea_95}, \cite{deO_16}.

\noindent{\bf Contribution in the context of current Literature.} While being directly inspired by the current Literature on network de-synchronization, this work offers a number of key novelties: 
\begin{itemize}
\item this paper considers network dynamics which are globally coupled via a quorum sensing (global, or shared) variable. With respect to this, the key novelty is that it considers the global variable having its own dynamics, modeled via a set of ODEs. Such a dynamics, in turn, depends on the quorum variable and on the state variables of the network nodes (also modeled via ODEs);
\item a sufficient condition is provided for de-synchronization in quorum-sensing networks;
\item finally, this paper also illustrates via two applications how the results can be effectively used to predict the onset of de-synchronization.
\end{itemize}

The paper is organized as follows. We start in  Section \ref{sec:model_statement} with defining the models considered in this paper and formalizing the problem statement. In Section \ref{sec:lemma} we give two new lemmas which are then used in Section \ref{sec:desynch} to devise our main result on the de-synchronization of quorum-sensing networks. The effectiveness of our approach is shown in Section \ref{sec:applications}, where we use our results to study de-synchronization in networks from two motivating applications. Concluding remarks are offered in Section \ref{sec:conclusions}. Finally, for the reader's convenience, the key mathematical tools used to prove our results are given in the Appendix.

\section{Mathematical formulation and problem statement}\label{sec:model_statement}
The goal of this Section is to introduce the networks considered in this paper and to give a definition for network de-synchronization. Such a definition is based on the concept of trajectories divergence.

\subsection{Trajectories divergence}
\begin{figure}[thbp]
\begin{center}
  \includegraphics[width=8cm]{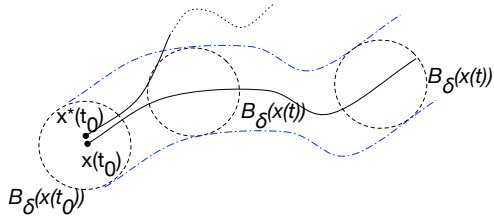}
  \caption{Geometric interpretation of Definition \ref{def:basic_divergence}. Two trajectories, $x(t)$ (with initial condition $x(t_0)$) and $x^{\ast}(t)$ (with initial condition $x^{\ast}(t_0)$) are shown. The open sets $\mathcal{B}_{\delta}(x(t))$ define, over  time, an open bundle (in blue in the figure, colors on-line) around $x(t)$. The two trajectories have nearby initial conditions, i.e. $x^{\ast}(t_0)$ belongs to $\mathcal{B}_{\delta}(x(t_0))$. The distance between trajectory $x(t)$ and $x^{\ast}(t)$ increases and this causes $x^{\ast}(t)$ to  exit from the bundle defined by $\mathcal{B}_{\delta}(x(t))$. Note that Definition \ref{def:basic_divergence} does not provide any insight on how the distance $\abs{x^{\ast}(t) - x(t)}$ evolves once $x^{\ast}(t)$ is outside of the bundle.}  
  \label{fig:basic_divergence}
  \end{center}
\end{figure}

We now formalize the notion of divergence between two solutions (or trajectories) for the generic nonlinear dynamical system (\ref{eqn:gensys}). In order to do so, let $x(t)$ be a solution of (\ref{eqn:gensys}) and assume that the solution exists for $\forall t\ge t_0$. Then, we denote by $\mathcal{B}_{\delta}(x(t))$ some open ball (or neighborhood) of radius $\delta > 0$ around $x(t)$ at time $t$. We are now ready to give the following definition.
\begin{Definition}\label{def:basic_divergence}
Let $x(t)$ and $x^{\ast}(t)$ be two different solutions of (\ref{eqn:gensys}), with $x^{\ast}(t_0) \in \mathcal{B}_{\delta}(x(t_0))$. We say that $x^{\ast}(t)$ is diverging with respect to $x(t)$ if there exists some $K \ne 0$ and some $d \ne 0$ such that $\abs{x^{\ast}(t) - x(t)} \ge \bar K^2 e^{d^2(t-t_0)}$, $\forall t$ such that $x^{\ast}(t) \in \mathcal{B}_{\delta}(x(t))$.
\end{Definition}
In the rest of the paper, we will simply say that the dynamics (\ref{eqn:gensys}) is diverging with respect to $x(t)$ if the above definition is fulfilled for all the trajectories $x^\ast(t)$ such that $x^\ast(t_0) \in \mathcal{B}_{\delta}(x(t_0))$. We now offer the following remarks:
\begin{itemize}
\item the set $\mathcal{B}_{\delta}(x(t))$ defines, over time, an open bundle around the trajectory $x(t)$;
\item a geometric interpretation of Definition \ref{def:basic_divergence} is given in Figure \ref{fig:basic_divergence}. In such a figure, two neighboring trajectories are shown, i.e. $x(t)$ and $ x^{\ast}(t)$, with $x^{\ast}(t)$ diverging with respect to $x(t)$.
\end{itemize}

\subsection{Network model and de-synchronization}
\begin{figure}[thbp]
\begin{center}
  \includegraphics[width=6cm]{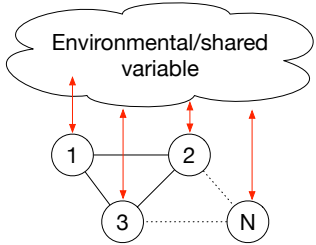}
  \caption{Networks considered in this paper. Network nodes interact with each other and with a shared environmental variable. Both network nodes and the shared variable are modeled via a set of ODEs.}
  \label{fig:quorum_physics}
  \end{center}
\end{figure}

Throughout this paper, we will consider networks where a set of agents, modeled via a set of smooth ordinary differential equations, communicates with each other. In addition to this direct node-to-node link, nodes also communicate indirectly, through a shared (environmental) variable, which is also modeled by a set of ODEs. The structure of these networks is schematically shown in Figure \ref{fig:quorum_physics}. For the applications of interest in this paper and discussed in Section \ref{sec:applications}, the shared variable will either be a service with which network nodes interact or a shared molecule concentration surrounding certain biochemical entities.

Formally, the networks that we will consider will be described with the following smooth differential equation:
\begin{equation}\label{eqn:net}
\begin{array}{*{20}l}
\dot x_i = f(t,x_i) + \Gamma(t) u\left(\sum_{j \in \mathcal{N}_i} \left(g(x_j) - g(x_i)\right)\right) + h_x(t,x_i,z), \\
\dot z = r(t,z) + h_z(t,z,X),
\end{array}
\end{equation}
$\forall t \ge t_0$, $t_0\ge 0$ where: (i) $x_i \in \R^n$ is the state variable for the $i$-th network node and $i = 1\ldots,N$; (ii) $X(t) = [x_1^T,\ldots,x_N^T]^T$ is the stack of the nodes state variables and $X(t_0) := X_0$; (iii) $f(\cdot,\cdot):\R^+\times\R^n\rightarrow\R^n$ models the nodes intrinsic dynamics; (iv) $z \in \R^m$ is the shared variable with which all network nodes interact, $z(t_0):= z_0$ and $r(\cdot,\cdot): \R^+\times \R^m\rightarrow\R^m$ models the intrinsic dynamics of such a variable; (v) $h_x(\cdot,\cdot,\cdot):\R^+\times\R^n\times\R^m\rightarrow\R^n$ and $h_z(\cdot,\cdot,\cdot):\R^+\times\R^m\times\R^{nN}\rightarrow\R^m$ model the interaction between network nodes and the shared variable; (vi) $u(\cdot):\R^n\rightarrow\R^n$ is a smooth function describing the direct coupling between nodes; (vii) $\Gamma(t)$ is an $n\times n$ time varying function modeling the coupling strength; (viii) the function $g(\cdot):\R^n\rightarrow\R^n$ is a smooth output function for network nodes; (ix) $\mathcal{N}_i$ is the set of neighbors to node $i$.

In the rest of this paper we assume that, for some $x_s(t) \in\R^n$, a solution of the form $\tilde S(t) = [S(t)^T,z(t)^T]^T$, $S(t) := 1_N\otimes x_s(t)$, exists for network (\ref{eqn:net}). The solution $\tilde S(t)$ is characterized by the fact that all the network nodes evolve onto the same trajectory, $x_s(t)$. For this reason, we will say that $\tilde S(t)$ is the synchronous solution of (\ref{eqn:net}). The goal of this paper is to provide a sufficient condition for network de-synchronization.  This can be formalized in terms of divergence of the network trajectories with respect to $S(t)$, i.e. with respect to a component of $\tilde S(t)$.
\begin{Definition}\label{def:diverging_sys}
We say that (\ref{eqn:net}) de-synchronizes if there exists at least one  dynamics transversal to the synchronization manifold which is diverging with respect to $S(t)$. 
\end{Definition}

Intuitively, Definition \ref{def:diverging_sys} implies that all the solutions of (\ref{eqn:net}) starting close to the synchronization manifold locally diverge from the synchronous solution. This will be useful for proving Theorem \ref{thm:net_desynch}, when we will prove de-synchronization by showing that at least one eigendirection transversal to the synchronization manifold is diverging.

In the rest of the paper, we will simply say that (\ref{eqn:net}) is de-synchronizing if it fulfills Definition \ref{def:diverging_sys}. Please note that the property given in Definition \ref{def:diverging_sys} is a local differential property as it is defined for all the trajectories which are sufficiently close to the solution of interest. Note also that the definition involves only the trajectories of the network nodes ($x_i$'s), without specifying the behavior of the environmental variable, $z(t)$.

\section{Diverging lemmas}\label{sec:lemma}
We now introduce two lemmas that will be used in Section \ref{sec:desynch} to prove the main result of this paper. The lemmas make use of the concept of matrix measure, $\mu$, which is formally introduced in the appendix.

With the Lemma below we provide a sufficient condition for (\ref{eqn:gensys}) to be diverging with respect to some desired solution, say $x_d(t)$.
\begin{Lemma}\label{lem:div_general}
Assume that for system (\ref{eqn:gensys}), there exists some matrix measure and some $d\ne 0$ such that
$$
\mu\left(-\frac{\partial f}{\partial x}(t,x_d)\right) \le - d^2, 
$$ 
$\forall t \in\R^+$. Then, (\ref{eqn:gensys}) is diverging with respect to $x_d(t)$.
\end{Lemma}
\proof See the Appendix. \endproof

With the next Lemma, we will instead consider a dynamical system composed by two interconnected subsystems (say subsystem $a$ and subsystem $b$) described by the following smooth differential equation:
\begin{equation}\label{eqn:struc_sys}
\begin{array}{*{20}l}
\dot p = a(t,p,q),\\
\dot q = b(t,q,p),\\
\end{array}
\end{equation}
where $a(\cdot,\cdot,\cdot):\R^+\times\R^n\times\R^m\rightarrow\R^n$ and $b(\cdot,\cdot,\cdot):\R^+\times\R^m\times\R^n\rightarrow\R^m$. Let $[p_d(t)^T, q_d(t)^T]^T$ be the desired solution for (\ref{eqn:struc_sys}). The following result provides a sufficient condition for the divergence of subsystem $a$ with respect to $p_d(t)$.

\begin{Lemma}\label{lem:div}
Consider system (\ref{eqn:struc_sys}) and let $q^\ast(t)$ be the solution of $\dot q^\ast(t) = b(t,q^\ast,p_d)$. Then, subsystem $a$ is diverging with respect to $p_d$
if the reduced-order auxiliary system 
$$
\dot y_p = a(t,y_p,q^\ast(t)),
$$
is diverging with respect to $p_d(t)$.
\end{Lemma}
\proof See the Appendix. \endproof

We remark that, in Lemma \ref{lem:div_general}, $\frac{\partial f}{\partial x}$ is the $n\times n$ Jacobian matrix of the vector field of system (\ref{eqn:gensys}), i.e. $f(t,x)$. Therefore, such a Lemma is essentially a condition on the matrix measure of the Jacobian of system (\ref{eqn:gensys}).

\section{De-synchronization in quorum-sensing networks}\label{sec:desynch}
We are now ready to state the main result of the paper, which provides a sufficient condition for the de-synchronization of (\ref{eqn:net}).
\begin{Theorem}\label{thm:net_desynch}
Assume that for (\ref{eqn:net}) there exists a matrix measure, $\mu$, some $d\ne 0$ and some $i$, $2\le i\le N$ such that:
\begin{equation}\label{eqn:neur_cond}
\lambda_i\mu\left(\Gamma(t)\frac{\partial u}{\partial x}(0)\frac{\partial g}{\partial x}(x_s)\right) + \mu\left(-\frac{\partial f}{\partial x}(t,x_s)-\frac{\partial h_x}{\partial x}(t,x_s,z)\right)\le -d^2,
\end{equation}
$\forall z\in\R^m$. Then, (\ref{eqn:net}) de-synchronizes.
\end{Theorem}
\proof We will prove de-synchronization by proving that there exists at least one diverging eigendirection transversal to the synchronization manifold. Following Lemma \ref{lem:div}, de-synchronization can be proved by proving de-synchronization of the following reduced order  auxiliary system
\begin{equation}\label{eqn:net_virtual_2}
\dot y_i = f(t,y_i) + \Gamma(t) u\left(\sum_{j \in \mathcal{N}_i} \left(g(y_j) - g(y_i)\right)\right) + h_x(t,y_i,z(t)).
\end{equation}
Note that the synchronous solution of (\ref{eqn:net}) is also a solution of (\ref{eqn:net_virtual_2}). We will prove de-synchronization by proving that for network (\ref{eqn:net_virtual_2}) there exists at least one diverging eigendirection transversal to the synchronization manifold. Linearizing the dynamics (\ref{eqn:net_virtual_2}) around the synchronous trajectory yields:
\beqn
\dot \delta y_i = \frac{\partial f}{\partial y}\left(t,x_s\right)\partial \delta y_i 
 + \Gamma(t) \frac{\partial u}{\partial y}(0)\sum_{j\in N_i}\left(\frac{\partial g}{\partial y}(x_s)\delta y_j 
-
\frac{\partial g}{\partial y}(x_s)\delta y_i\right) + \frac{\partial h_x}{\partial y}(t,x_s,z(t))\delta y_i,
\eeqn
where $\delta y_i = y_i -x_s(t)$. Now, let $\delta Y := [\delta y_1^T,\ldots,\delta y_N^T]^T$, we can then rewrite the whole network dynamics as 
\begin{equation}\label{eqn:error_lin}
\delta \dot Y = \left(I_N \otimes \left(\frac{\partial f}{\partial y}\left(t,x_s\right)+\frac{\partial h_x}{\partial y}(t,x_s,z)\right)\right)\delta Y  - \left(L\otimes \Gamma(t)\frac{\partial u}{\partial y}(0)\frac{\partial g}{\partial y}(x_s)\right)\delta Y .
\end{equation}
Since the network topology is undirected, we have that $L$ is symmetric. Therefore, by means of Lemma \ref{lem:schur} {\bf (see the Appendix)} we have that there exists an $N\times N$ orthogonal matrix $Q$ ($Q^TQ=I_N$) such that $\Lambda=Q^{T}LQ$, where $\Lambda$ is the $N\times N$ diagonal matrix, having on its main diagonal the eigenvalues of $L$. Define the  coordinate transformation $\delta Y^\ast = \left(Q \otimes I_n\right)^{-1} \delta Y$. In the new coordinates, (\ref{eqn:error_lin}) becomes
$$
\begin{aligned}
\delta \dot Y^\ast= (Q\otimes I_n)^{-1} \left[I_N\otimes \left(\frac{\partial f}{\partial y}\left(t,x_s\right) +\frac{\partial h_x}{\partial y}(t,x_s,z(t))\right) \right. + \\
\left. -\left(L\otimes \Gamma(t)\frac{\partial u}{\partial y}(0)\frac{\partial g}{\partial y}(x_s)\right)\right] (Q\otimes I_n) \delta Y^\ast,
\end{aligned}
$$
which can be written as:
\begin{equation}\label{eqn:net_lin_transf}
\delta \dot Y^\ast = \left[I_N \otimes \left(\frac{\partial f}{\partial y}\left(t,x_s\right) +\frac{\partial h_x}{\partial y}(t,x_s,z(t))\right)  \right.
-\left.\Lambda \otimes \Gamma (t)\frac{\partial u}{\partial y}(0)\frac{\partial g}{\partial y}(x_s)\right] \delta Y^\ast,
\end{equation}
or, equivalently:
\begin{equation}\label{eqn:transv_dyn}
\delta \dot y_i^\ast  =  \left[\left(\frac{\partial f}{\partial y}\left(t,x_s\right) \right.+ \frac{\partial h_x}{\partial y}(t,x_s,z(t))\right)
 - \left. \lambda_i\Gamma(t)\frac{\partial u}{\partial y}(0)\frac{\partial g}{\partial y}(x_s)\right] \delta y_i^\ast,
\end{equation}
$i = 1,\ldots, N$, $y_i^\ast \in \R^n$ and where Lemma \ref{lem:kronecker} has been used {\bf (see the Appendix)}. Indeed, by means of such a result we have $\left(Q\otimes I_n\right)^{-1} \left(I_N\otimes \left(\frac{\partial f}{\partial y}\left(t,x_s\right) +\frac{\partial h_x}{\partial y}(t,x_s,z(t)) \right)\right) \left(Q\otimes I_n\right) = \left(I_N\otimes \left(\frac{\partial f}{\partial y}\left(t,x_s\right)+\frac{\partial h_x}{\partial y}(t,x_s,z(t)))\right) \right)$ and $\left(Q\otimes I_n\right)^{-1} \left(L \otimes \Gamma(t)\frac{\partial u}{\partial y}(0)\frac{\partial g}{\partial y}(x_s)\right)\left(Q\otimes I_n\right)
=\Lambda\otimes \Gamma(t)\frac{\partial u}{\partial y}(0)\frac{\partial g}{\partial y}(x_s)$.

Now, the network de-synchronizes if at least one of the  dynamics transversal to the synchronization manifold is diverging. In turn, the dynamics transversal to such a subspace are those in (\ref{eqn:transv_dyn}) with $i=2,\ldots,N$. That is, following Lemma \ref{lem:div_general}, the network is diverging if for some $i$, $2\le i \le N$, it happens that
\beqn
\mu\left(-\frac{\partial f}{\partial y}\left(t,x_s\right) - \frac{\partial h_x}{\partial y}(t,x_s,z(t)) \right. 
\left.+ \lambda_i \Gamma(t)\frac{\partial u}{\partial y}(0)\frac{\partial g}{\partial y}(x_s)\right)\le -d^2
\eeqn

Since $\lambda_i$'s are positive for all $i= 2,\ldots,N$ we have \cite{Vid_93}:
$$
\begin{aligned}
\mu\left(-\frac{\partial f}{\partial y}\left(t,x_s\right) - \frac{\partial h_x}{\partial y}(t,x_s,z(t))+ \lambda_i \Gamma(t) \frac{\partial u}{\partial y}(0)\frac{\partial g}{\partial y}(x_s)\right) \le \\
\mu\left(-\frac{\partial f}{\partial y}\left(t,x_s\right)-\frac{\partial h_x}{\partial y}(t,x_s,z(t))\right) + \lambda_i\mu\left(\Gamma(t) \frac{\partial u}{\partial y}(0)\frac{\partial g}{\partial y}(x_s)\right).
\end{aligned}
$$
The proof is then concluded by noticing that, by hypotheses, at least one of the dynamics transversal to the synchronization manifold is diverging. This proves the result.
\endproof

\section{Applications}\label{sec:applications}

\subsection{When distributed sensing cannot be trusted}\label{sec:distr_sensing}
\begin{figure}[thbp]
\begin{center}
  \includegraphics[width=10cm]{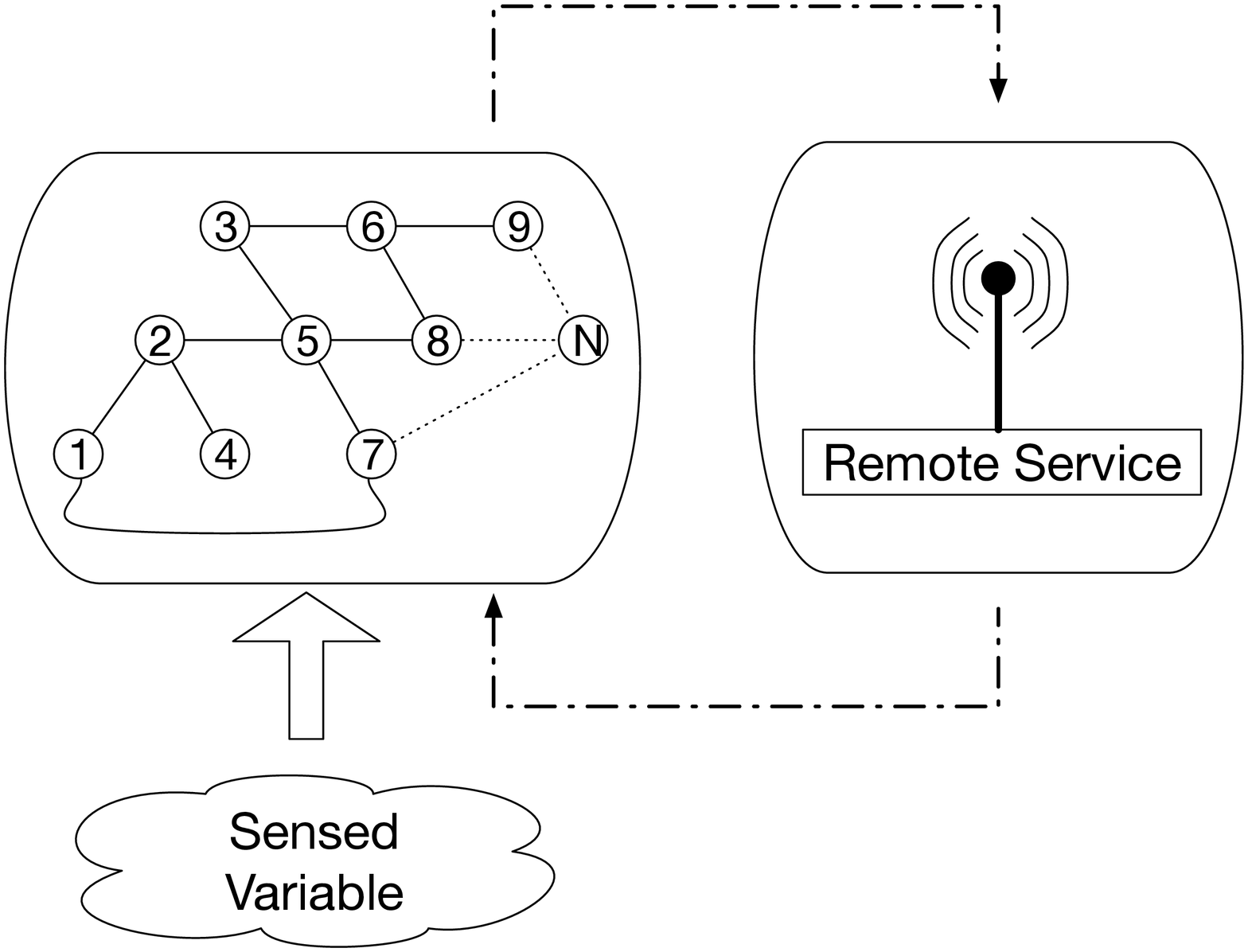}
  \caption{A motivating application, where a sensor network estimates a quantity of interest and sends the estimate to a remote service.}
  \label{fig:distributed_sensing}
  \end{center}
\end{figure}

The so-called Internet of Things (IoT) revolution is allowing us to connect objects in ways that were not even imaginable a few years ago. This is leading to interesting applications for smart cities as it gives the possibility of creating pervasive networks of actuators/sensors deployed in urban environments. The goal of such networks is typically that of monitoring a given quantity of interest (e.g. air quality, gas leakages, weather, ...), gather some aggregate information from field data and send this information to base stations. Here, the aggregate data are further analyzed in order to provide new smarter user services. The set-up outlined here, is schematically shown in Figure \ref{fig:distributed_sensing}, where a network consisting of $N$ devices is deployed to the field in order to sense some distributed quantity. The aggregate information is then sent to a base station which performs additional filtering, forwards these data to analytics algorithms and provides feedback to the devices. Our motivating question is then: {\em when can we trust the information provided by the network?}
 
The network in Figure \ref{fig:distributed_sensing} can be modeled as a quorum-sensing network, where: (i) the IoT devices deployed to the field are the network nodes; (ii) the base station has the role of the shared environment. In this Section, we will consider the following network:
\begin{equation}\label{eqn:sensing_net}
\begin{array}{*{20}l}
\dot x_i & = &  \bar q(t) - x_i + \gamma_1(t)\sum_{j\in N_i}\left(k\left(x_j^3-x_j\right)- k\left(x_i^3-x_i\right)\right) + \gamma_2(t)\left(z-x_i\right),\\
\dot z & = & - z + \frac{1}{N}\sum_{i=1}^N(x_i-z)
\end{array}
\end{equation}
where $i=1,\ldots,N$, $x_i\in\R$, $\bar q(t)$ is the quantity that is being sensed by the network of devices. In (\ref{eqn:sensing_net}), $\gamma_1(t)$ and $\gamma_2(t)$ are respectively the time varying node-to-node and node-to-base-station coupling strengths, while $k$ is the gain of the coupling protocol between nodes. 
 
Please note that (\ref{eqn:sensing_net}) can be recast onto (\ref{eqn:net}) with $f(t,x) := \bar q(t)  - x$, $\Gamma(t) :=\gamma_1(t)$ and $u(x) := x$, $g(x) := k\left(x^3-x\right)$, $h_x(t,x,z) := \gamma_2(t)(z-x)$, $r(t,z) := - z$, $h_z(t,z,X) := 1/N\sum_{i=1}^N(x_i-z)$. The task for which the network is designed is to ensure that all nodes will sense $x_d := \bar q(t)$, i.e. that nodes converge towards the solution $X_d = 1_N\otimes \bar q(t)$. We will now use Theorem \ref{thm:net_desynch} to obtain a straightforward sufficient condition ensuring that the network will be diverging with respect to $X_d$. Following Theorem \ref{thm:net_desynch}, de-synchronization can be characterized in terms of the network algebraic connectivity, $\lambda_2$. Specifically, the condition of Theorem \ref{thm:net_desynch} with $i=2$ implies that the network de-synchronizes if there exists some matrix measure, $\mu$, such that
$$
\lambda_2\mu\left(k\gamma_1(t)\left(3\bar q(t)^2 - 1)\right)\right) \le -\mu\left(1+\gamma_2(t)\right).
$$ 

Since network nodes are $1$-dimensional, this translates to:
\begin{equation}\label{eqn:sensing_divergence_cond}
\lambda_2k\gamma_1(t)\left(3\bar q(t)^2-1)\right) \le -1-\gamma_2(t).
\end{equation}

\begin{figure}[thbp]
\begin{center}
  \includegraphics[width=8cm]{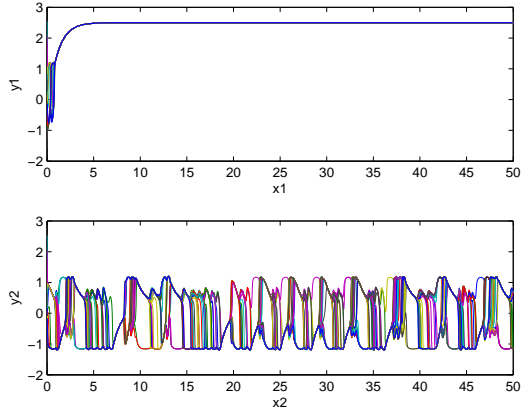}
  \caption{Time evolution for the small world network considered in Section \ref{sec:distr_sensing}. Time is on the $x$-axis and $x_i$'s are on the $y$-axis. In the top panel the network nodes' behavior is shown for $\bar q = 2.5$: such a panel shows that all the nodes properly sense the quantity of interest. In the right panel, the nodes' time behavior is instead shown for $\bar q = 0.25$. In such a panel, nodes are not able to sense the quantity of interest as no agreement is reached. Initial conditions for the network nodes are taken from a standard distribution.}
  \label{fig:net_sim}
 \end{center}
\end{figure}

That is, if the above condition  occurs, then the network will be diverging with respect to $X_d$, thus implying that the network will no longer properly sense $\bar q(t)$. Now, (\ref{eqn:sensing_divergence_cond}) provides an explicit condition on the node-to-node communication network topology (via $\lambda_2$) and coupling design (via $\gamma_1(t)$, $\gamma_2(t)$ and $k$). Specifically, if network topology and coupling are not well blended together, then the network will not properly sense $\bar q_d$, i.e. it will not perform the task for which it has been designed. Also, please note that the higher the $\gamma_2$, then the more difficult will be to fulfill the condition in (\ref{eqn:sensing_divergence_cond}), thus helping to prevent network de-synchronization. Assume that $k = \gamma_1 = \gamma_2 =1$. As a testbed network, we consider a small world network of $N = 50$ nodes generated by following the method in \cite{Wat_Str_98}. We calculated numerically the eigenvalues of the Laplacian and found that, in this case, the algebraic connectivity for the network of our interest is $10$. Therefore, our condition for de-synchronization becomes: $10(3\bar q(t)^2-1) \le -2$. This means that if the quantity of interest $\bar q(t)$ becomes too small, then the network will not be able to properly sense it. This prediction is confirmed in Figure \ref{fig:net_sim}.

\subsection{De-synchronization of biochemical networks}\label{sec:FN} 
Over the last few years, synchronization of biochemical systems has attracted much research efforts both from the theoretical, see e.g. \cite{Str_03} and experimental \cite{Yag_Ise_Mat_Oku_Yag_03} viewpoints. Specifically, the importance of synchronization for such networks has motivated a large body of results aimed at providing sufficient conditions for network synchronization (see e.g. \cite{Rus_diB_09b}, \cite{Rus_diB_09} and references therein). We now address the following motivating question: {\em given a synchronized biochemical network of interest, which are the mechanisms that lead to the loss of synchronization?} This is a relevant question for a large number of biochemical applications, with a remarkable example being the fact that de-synchronization is believed to be an indicator of metabolic diseases (see e.g. \cite{Ahn_15}, \cite{Rip_12}). We now consider the following network: 
\begin{equation}\label{eqn:FN_net}
\begin{array}{*{20}l}
\dot x_i & = & -\delta x_i +k_1y_i-k_2\left(E_T-y_i\right)x_i + \gamma_1 (t)  u\left(\sum_{j\in\mathcal{N}_i}\left(x_i-x_j\right)\right) +\gamma_2(t)\frac{K_1z}{K_2+z} \\
\dot y_i & = & -k_1y_i+k_2\left(E_T-y_i\right)x_i\\
\dot z & = & -\sum_{i=1}^N\frac{K_1z}{K_2 + z} + i(t),
\end{array}
\end{equation}
where in this case the shared environmental variable models a biochemical reaction between a set of $N > 1$ enzymes sharing the same substrate (see e.g. \cite{Sza_Ste_Per_06}). The nodes' dynamics in (\ref{eqn:FN_net}) are particularly relevant in systems and synthetic biology as it models a general externally-driven transcriptional module. Such transcriptional modules are ubiquitous in biology, natural as well as
synthetic, and their behavior was recently studied in~\cite{DelV_Nin_Son_08}
in the context of ``retroactivity'' (impedance or load) effects. The state variables $x_i$'s are the concentrations of  generic transcription factors, say ($X_i$'s). The state variables $y_i$'s are the concentrations of complex proteins-promoters, say $Y_i$'s. The production of each $y_i$ is stimulated by the corresponding $x_i$. The time evolution of the substrate is modeled by the dynamics of $z(t)$ and its production is stimulated by a time dependent input function $i(t)$, which is a positive function. Please refer to \cite{DelV_Nin_Son_08} for a detailed discussion on (\ref{eqn:FN_net}). In the same paper it is also shown that the quantities $E_t-y_i$ are always positive and that the system evolves on the positive orthant. In \cite{Rus_diB_Son_10}, the  transcription module has been analyzed to show that it can be entrained by any periodic input. Furthermore, in the same paper, the authors also proved that network (\ref{eqn:FN_net}) can be always synchronized if the coupling between nodes is linear and diffusive. Unfortunately, when modeling biochemical networks, it is often the case where the coupling is not linear and diffusive but it is rather a sigmoid function (modeling transcriptional interactions, see e.g. \cite{Sza_Ste_Per_06}). Motivated by this, we now investigate the effects on such a coupling function on the synchronization properties of the network. We will consider network nodes being coupled via a decreasing sigmoig function, i.e. $u(x) := 1/(1+e^x)$. We will then use Theorem  \ref{thm:net_desynch} to provide an effective sufficient condition to determine when the network will de-synchronize. 

We will now use again Theorem \ref{thm:net_desynch} to provide a sufficient condition for de-synchronization in terms of $\lambda_2$. The first step to apply Theorem \ref{thm:net_desynch} is to choose a matrix measure to verify (\ref{eqn:neur_cond}). In analogy to \cite{Rus_diB_Son_10}, in  what follows we will the matrix measure induced by the vector-$1$ norm, $\mu_1$.  In order to apply our result, first note that
$$
\mu_1\left(\Gamma(t)\frac{\partial u}{\partial x}(0)\right) = -\gamma_1(t)\frac{1}{4},
$$
while
$$
\begin{array}{*{20}l}
\mu_1\left(-\frac{\partial f}{\partial x}-\frac{\partial h_x}{\partial x}\right) & = & \mu_1\left(\left[\begin{array}{*{20}c}
\delta +k_2(E_T-y_i) & -(k_1+k_2x_i)\\
-k_2(E_T-y_i) & k_1+k_2x_i
\end{array}\right]\right) \\
& = & \max\left\{\delta+2k_2(E_T-y_i), 2(k_1+k_2x)\right\}.
\end{array}
$$

Due to the physical constraints of the system, we have $\delta+2k_2(E_T-y_i) \le \delta 2k_2E_T$ and $2(k_1+k_2x) \le 2(k_1+k_2\bar X)$, where $\bar X$ is the maximum of $x(t)$ (note that system trajectories are bounded if $i(t)$ is a bounded signal, see \cite{Rus_diB_Son_10}). Therefore:
$\mu_1\left(-\frac{\partial f}{\partial x}\right) \le \max\left\{\delta + 2k_2E_T, 2(k_1+k_2\bar X)\right\}$.

Thus, following Theorem \ref{thm:net_desynch}, the network will de-synchronize if 
$$
-\lambda_2\frac{\gamma_1(t)}{4} < - \max\left\{\delta +2k_2E_T, 2(k_1+k_2\bar X)\right\},
$$
i.e. if $\gamma_1$ and/or $\lambda_2$ become sufficiently large. Note that, in this case, the condition for de-synchronization does depend on $\gamma_2 (t)$. 

\begin{figure}[thbp]
\begin{center}
  \includegraphics[width=8cm]{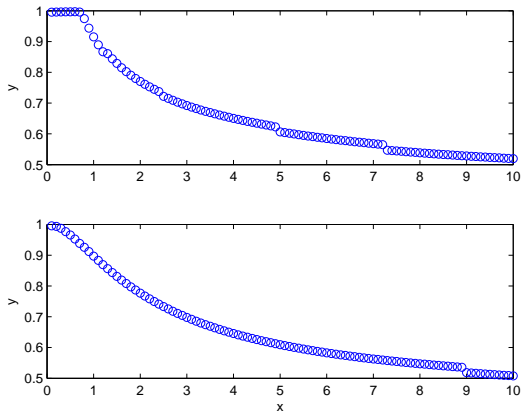}
  \caption{Order parameter as a function of $\gamma_1$. Values of $\gamma_1$ on the $x$-axis and the order parameter $R$ on the $y$-axis. The increase of this parameter causes a loss of synchronization for both Network $1$ (top panel) and Network $2$ (bottom panel). Note that Network $1$ starts to de-synchronize after Network $2$, thus confirming our theoretical predictions.}
  \label{fig:bio_sim_1}
 \end{center}
\end{figure}

\begin{figure}[thbp]
\begin{center}
  \includegraphics[width=8cm]{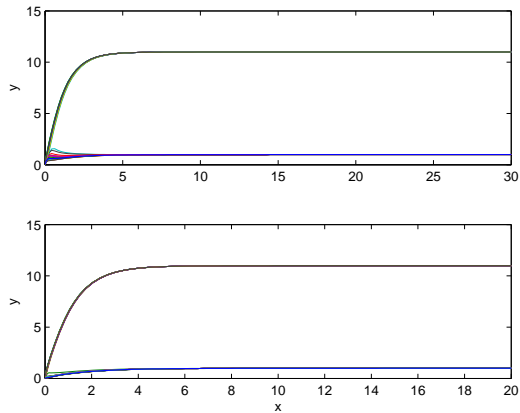}
  \caption{Time is on the $x$-axis and $x_i$'s on the $y$-axis. The time evolution for Network $1$ (top panel) and Network $2$ (bottom panel) is shown when $\gamma_1 = 10$. The two panels show that two groups (or clusters) of synchronized nodes emerge.}
  \label{fig:sim_bio_2}
  \end{center}
\end{figure}

In order to validate our theoretical prediction, we consider two small-world networks of $50$ nodes, say Network $1$ and Network $2$. The two networks are characterized by two different algebraic connectivity values ($\lambda_2 = 0.1$ for Network $1$ and $\lambda_2 = 13$ for Network $2$). The network parameters that we considered were: $i(t) = 1 +\sin (t)$, $k_1=E_T=\delta= K_1 = K_2 =1$, $k_2 = 0.1$, $\gamma_2 = 1$. In order to characterize quantitatively the  level of synchronization of the networks we used the {\em order parameter} 
$R := (\langle M^2\rangle - \langle M\rangle^2)/(\overline{\langle v_i^2\rangle - \langle v_i\rangle^2})$, defined following \cite{Gar_Elo_Str_04} where: (i) $M(t) := 1/N\sum_{i=1}^Nx_i$; (ii) $\langle \cdot \rangle$ denotes the time average; (ii) $\bar{\cdot}$ denotes the average over the network nodes. In Figure \ref{fig:bio_sim_1} the order parameter is plotted as a function of $\gamma_1$ for both Network $1$ and Network $2$. As shown in such a figure, the increase in $\gamma_1$ causes a network transition from a synchronized state towards an un-synchronized state. Moreover, as expected from our theoretical predictions, Network $1$ starts to de-synchronize after Network $2$. Essentially, this is due to the fact that Network $2$ has a larger algebraic connectivity than Network $1$.  Finally, in Figure \ref{fig:sim_bio_2} the networks behavior is shown when $\gamma_1 = 10$. As shown in such a figure, the increase in $\gamma_1$ causes a loss of network synchronization. In particular, two separate groups (or clusters) of nodes emerge, with each group being synchronized onto a different trajectory. The emergence of why this phenomenon happens will be subject of future research.

\section{Conclusions}\label{sec:conclusions}
In this paper we presented a sufficient condition for the de-synchronization of quorum-sensing networks. After presenting our main result, we showed the effectiveness of our approach by considering two networks arising in the contexts of distributed sensing and biochemical networks. In presenting new conditions for network de-synchronization, our work also opens new questions. Of particular interest is the understanding of why, for some specific dynamics like those arising in biology, de-synchronization leads to clustering effects where two or more clusters of synchronous nodes emerge. 

\section*{Acknowledgements}
The author wishes to acknowledge the Associate Editor and the anonymous reviewers for their invaluable comments and suggestions, which considerably improved the quality and the clarity of the paper.

\section*{Appendix}
\section*{Mathematical tools}\label{sec:preliminaries}
In this Section we introduce the notation, definitions and matrix properties that will be used in the rest of the paper. This Section also provides an introduction to concepts related to graphs and Laplacian matrices, which will be used in the paper.

\subsection*{Matrix notation and properties}

In this paper, $1_N$ will denote the $N$ dimensional column vector having all elements equal to $1$ and $I_N$ will denote the $N\times N$ identity matrix. Finally, $\otimes$ will be used to denote the Kronecker (or direct) product. The following two technical results will be useful in the rest of the paper (see e.g. \cite{Arc_10}).

\begin{Lemma}\label{lem:kronecker}
The following properties hold for the Kronecker product: (i) $\left(A \otimes B \right) \left(C \otimes D\right) = \left(A C\right) \otimes\left(BD\right)$; (ii) if $A$ and $B$ are invertible, then $\left(A \otimes B \right)^{-1} =A^{-1}\otimes B^{-1}$.
\end{Lemma}
\begin{Lemma}\label{lem:schur}
For any $n\times n$ real symmetric matrix, $A$, there exist an orthogonal $n\times n$ matrix, $Q$, such that $Q^TAQ= U$, where $U$ is an $n\times n$ diagonal matrix.
\end{Lemma}

\subsection*{Matrix measures}
We  recall (see for instance~\cite{michelbook}) that, given a vector norm on Euclidean space ($\abs{\cdot}$), with its induced matrix norm $\norm{A}$, the associated matrix measure (or  logarithmic norm, see \cite{dahlquist,lozinskii}) $\mu$ is defined as 
$
\mu(A) \,:=\;
\lim_{h \rightarrow 0^+} \frac{1}{h} \left(\norm{I+hA}-1\right)$.
The above limit is known to exist, and the convergence is monotonic,
see~\cite{strom,dahlquist}. Some matrix measures are reported in Table \ref{tab:matrix_measures}.

\begin{table}[th] 
\caption{Common matrix measures for a real $n\times n$ matrix, $A=[a_{ij}]$. The $i$-th eigenvalue of $A$ is denoted with $\lambda_i(A)$.}
\centering
\label{tab:matrix_measures}
\begin{tabular}{|c| c|} 
\hline
vector norm, $\abs{\cdot}$ & induced matrix measure, $\mu\left(A\right)$\\
\hline
$\abs{x}_1= \sum_{j=1}^n\abs{x_j}$ & $ \mu_1\left(A\right)= \max_{j} \left( a_{jj}+\sum_{i \ne j}  \abs{a_{ij}} \right)$\\
\hline
$\abs{x}_2= \left( \sum_{j=1}^n\abs{x_j}^2\right)^{\frac{1}{2}}$ & $ \mu_2\left(A\right)=\max_{i} \left( \lambda_i\left\{\frac{A+A^T}{2}\right\}\right)$\\
\hline
$\abs{x}_\infty= \max_{1 \le j \le n} \abs{x_j}$ & $\mu_{\infty}\left(A\right)= \max_{i} \left( a_{ii}+\sum_{j \ne i} \mid a_{ij}\mid \right)$\\
\hline
\end{tabular}
\end{table}

Recently, matrix measures have been used to devise upper bounds for the distances between trajectories of a dynamical system of interest. Specifically, let 
\begin{equation}\label{eqn:gensys}
\dot x = f \left(t,x \right),  \ \ \ x(t_0) = x_0, \quad t_0 \ge 0,
\end{equation}
be a smooth $n$-dimensional dynamical system evolving onto $\R^n$, with $J(t,x)$ being the system Jacobian. Then, as shown in \cite{Loh_Slo_98,Rus_diB_Son_10}, trajectories of  (\ref{eqn:gensys}) globally exponentially converge towards each other if there exists a matrix measure, $\mu$, such that $\mu(J(t,x))$ is uniformly negative. This approach is known as contraction analysis and it has been recently extended to the case of Caratheodory systems \cite{diB_Liu_Rus_14}. Contraction principles in metric functional spaces can be traced back to Banach and Caccioppoli (see e.g. \cite{Gra_03} for further details). In the field of continuous-time dynamical systems theory, ideas closely related to contraction can be found in \cite{Hartmann} and \cite{Lewis}. See also \cite{Pav_Pog_Wou_Nij}, ~\cite{Ang_02}, \cite{pde} and \cite{jou_05} for an historical overview. Recent results for the synchronization of complex networks via contraction can be instead found in \cite{Ami_Son_14}, while \cite{Ami_Son_14a} identifies some open problems of contraction methods for nonlinear systems.

\subsection*{Graphs}
We now revise some key notions from graph theory that will be used in this paper, \cite{Hor_Joh_99}. Let $\mathcal{G} := \left\{\mathcal{V},\mathcal{E}\right\}$ be an undirected graph, where $\mathcal{V}$ is the set of $N>1$ vertices (or nodes) and $\mathcal{E}\subseteq \mathcal{V} \times \mathcal{V}$ is the set of edges. We denote by $\mathcal{N}_i$ the set of neighbours to the $i$-th network node and we let $d_i$ be the number of its neighbours (i.e. $d_i$, also known as degree of node $i$, is the cardinality of $\mathcal{N}_i$). We will denote by $A$ the $N\times N$ graph adjacency matrix: the element $a_{ij}$ of $A$ is equal to $1$ if nodes $i$ and $j$ are neighbours, $0$ otherwise. The graph Laplacian matrix, $L$ can then be defined as follows: $L = \Delta - A$, where $\Delta$ is the $N\times N$ matrix having $\Delta_{ii} = d_i$. If the graph is undirected then, by construction, $L$ is symmetric. Moreover, $L$ is a $0$ column/row sum matrix and hence it has at least one eigenvalue equal to $0$. It can be shown, see e.g. \cite{Hor_Joh_99}, that, if $\mathcal{G}$ is connected, then it only has one $0$ eigenvalue and this corresponds to the eigenvector $1_N$. In the rest of the paper we will denote by $\lambda_i$, $i=1,\ldots,N$, the eigenvalues of $L$. The second smallest eigenvalue, $\lambda_2$, is termed as algebraic connectivity and it is non-zero if and only if $\mathcal{G}$ is connected.

\section*{Proof of Lemma \ref{lem:div_general}}
Pick any solution $x(t) \in \mathcal{B}\left(x_d(t)\right)$ and consider the virtual displacement, say $\delta x$, between $x(t)$ and $x_d(t)$. Then, the following exact differential relation holds (see e.g. \cite{Arn_78}, \cite{Rus_diB_Son_10}, \cite{Loh_Slo_98}):
$$
\delta \dot x = \left(\frac{\partial f}{\partial x}\left(t,x_d\right)\right) \delta x .
$$
By Coppel's inequality (see e.g. \cite{Vid_93}) we have that
$$
\abs{\delta x} \ge \abs{\delta x_0} e^{\int_{t_0}^t\left(-\mu\left(-\frac{\partial f}{\partial x}\left(\tau,x_d\right)\right)d\tau\right)}.
$$
Therefore, by hypotheses we have
$$
\abs{\delta x} \ge \abs{\delta x_0} e^{\int_{t_0}^td^2d\tau} = \abs{\delta x_0} e^{d^2(t-t_0)}:=\bar K^2e^{d^2(t-t_0)},
$$
thus proving the result.
\section*{Proof of Lemma \ref{lem:div}}
In order to prove the Lemma, consider the following auxiliary system, which has been first introduced in \cite{Wan_Slo_05}:
$$
\begin{array}{*{20}l}
\dot y_p = a(t,y_p,q^\ast)\\
\dot q^\ast = b(t,q^\ast,p_d),
\end{array}
$$
and note that, as shown in \cite{Rus_diB_Son_10}, \cite{Rus_Slo_10}, the desired solution $[p_d(t)^T,q_d(t)^T]^T$ is a trajectory of  this auxiliary system (to see this, it suffices to substitute $y_p$ with $p_d$ in the dynamics above). Note also that, for the auxiliary system, $q^\ast(t)$ is an exogenous input to the dynamics of $y_p(t)$. Therefore, following \cite{Rus_Slo_10} the dynamics of $y_p$ can be studied by just considering the reduced order auxiliary system
$$
\dot y_p = a(t,y_p,q^\ast(t)).
$$
Note that, by hypotheses: (i) $p_d(t)$ is a particular solution of the reduced order auxiliary system; (ii) the reduced order auxiliary system is diverging with respect to $p_d(t)$. Therefore, we have
\begin{equation}\label{eqn:virtual_ineq}
\abs{y_p(t) - p_d(t)} \ge \bar K^2 e^{d^2(t-t_0)}.
\end{equation}
Finally, since the solutions of (\ref{eqn:struc_sys}) are particular solution of the reduced order auxiliary system, (\ref{eqn:virtual_ineq}) implies that
$$
\abs{p(t) - p_d(t)} \ge \bar K^2 e^{d^2(t-t_0)},
$$ 
thus proving the result. \endproof

\begin{thebibliography}{69}%
\makeatletter
\providecommand \@ifxundefined [1]{%
 \@ifx{#1\undefined}
}%
\providecommand \@ifnum [1]{%
 \ifnum #1\expandafter \@firstoftwo
 \else \expandafter \@secondoftwo
 \fi
}%
\providecommand \@ifx [1]{%
 \ifx #1\expandafter \@firstoftwo
 \else \expandafter \@secondoftwo
 \fi
}%
\providecommand \natexlab [1]{#1}%
\providecommand \enquote  [1]{``#1''}%
\providecommand \bibnamefont  [1]{#1}%
\providecommand \bibfnamefont [1]{#1}%
\providecommand \citenamefont [1]{#1}%
\providecommand \href@noop [0]{\@secondoftwo}%
\providecommand \href [0]{\begingroup \@sanitize@url \@href}%
\providecommand \@href[1]{\@@startlink{#1}\@@href}%
\providecommand \@@href[1]{\endgroup#1\@@endlink}%
\providecommand \@sanitize@url [0]{\catcode `\\12\catcode `\$12\catcode
  `\&12\catcode `\#12\catcode `\^12\catcode `\_12\catcode `\%12\relax}%
\providecommand \@@startlink[1]{}%
\providecommand \@@endlink[0]{}%
\providecommand \url  [0]{\begingroup\@sanitize@url \@url }%
\providecommand \@url [1]{\endgroup\@href {#1}{\urlprefix }}%
\providecommand \urlprefix  [0]{URL }%
\providecommand \Eprint [0]{\href }%
\providecommand \doibase [0]{http://dx.doi.org/}%
\providecommand \selectlanguage [0]{\@gobble}%
\providecommand \bibinfo  [0]{\@secondoftwo}%
\providecommand \bibfield  [0]{\@secondoftwo}%
\providecommand \translation [1]{[#1]}%
\providecommand \BibitemOpen [0]{}%
\providecommand \bibitemStop [0]{}%
\providecommand \bibitemNoStop [0]{.\EOS\space}%
\providecommand \EOS [0]{\spacefactor3000\relax}%
\providecommand \BibitemShut  [1]{\csname bibitem#1\endcsname}%
\let\auto@bib@innerbib\@empty
\bibitem [{\citenamefont {Iniguez}\ \emph {et~al.}(2009)\citenamefont
  {Iniguez}, \citenamefont {Kertesz}, \citenamefont {Kaski},\ and\
  \citenamefont {Barrio}}]{Ini_09}%
  \BibitemOpen
  \bibfield  {author} {\bibinfo {author} {\bibfnamefont {G.}~\bibnamefont
  {Iniguez}}, \bibinfo {author} {\bibfnamefont {J.}~\bibnamefont {Kertesz}},
  \bibinfo {author} {\bibfnamefont {K.~K.}\ \bibnamefont {Kaski}}, \ and\
  \bibinfo {author} {\bibfnamefont {R.~A.}\ \bibnamefont {Barrio}},\
  }\href@noop {} {\bibfield  {journal} {\bibinfo  {journal} {Physical Review
  E}\ }\textbf {\bibinfo {volume} {80}},\ \bibinfo {pages} {066119} (\bibinfo
  {year} {2009})}\BibitemShut {NoStop}%
\bibitem [{\citenamefont {Li}\ \emph {et~al.}(2006)\citenamefont {Li},
  \citenamefont {Zheng},\ and\ \citenamefont {Hui}}]{Li_06}%
  \BibitemOpen
  \bibfield  {author} {\bibinfo {author} {\bibfnamefont {P.}~\bibnamefont
  {Li}}, \bibinfo {author} {\bibfnamefont {D.}~\bibnamefont {Zheng}}, \ and\
  \bibinfo {author} {\bibfnamefont {P.}~\bibnamefont {Hui}},\ }\href@noop {}
  {\bibfield  {journal} {\bibinfo  {journal} {Physical Review E}\ }\textbf
  {\bibinfo {volume} {73}},\ \bibinfo {pages} {056128} (\bibinfo {year}
  {2006})}\BibitemShut {NoStop}%
\bibitem [{\citenamefont {Gonze}\ \emph {et~al.}(2005)\citenamefont {Gonze},
  \citenamefont {Bernard}, \citenamefont {Walterman}, \citenamefont {Kramer},\
  and\ \citenamefont {Herzerl}}]{Gon_Ber_Wal_Kra_Her_05}%
  \BibitemOpen
  \bibfield  {author} {\bibinfo {author} {\bibfnamefont {D.}~\bibnamefont
  {Gonze}}, \bibinfo {author} {\bibfnamefont {S.}~\bibnamefont {Bernard}},
  \bibinfo {author} {\bibfnamefont {C.}~\bibnamefont {Walterman}}, \bibinfo
  {author} {\bibfnamefont {A.}~\bibnamefont {Kramer}}, \ and\ \bibinfo {author}
  {\bibfnamefont {H.}~\bibnamefont {Herzerl}},\ }\href@noop {} {\bibfield
  {journal} {\bibinfo  {journal} {Biophyscal Journal}\ }\textbf {\bibinfo
  {volume} {89}},\ \bibinfo {pages} {120} (\bibinfo {year} {2005})}\BibitemShut
  {NoStop}%
\bibitem [{\citenamefont {Anastassiou}\ \emph {et~al.}(2010)\citenamefont
  {Anastassiou}, \citenamefont {Montgomery}, \citenamefont {Barahona},
  \citenamefont {Buzsaki},\ and\ \citenamefont
  {Koch}}]{Ana_Mon_Bar_Buz_Koc_10}%
  \BibitemOpen
  \bibfield  {author} {\bibinfo {author} {\bibfnamefont {C.}~\bibnamefont
  {Anastassiou}}, \bibinfo {author} {\bibfnamefont {S.~M.}\ \bibnamefont
  {Montgomery}}, \bibinfo {author} {\bibfnamefont {M.}~\bibnamefont
  {Barahona}}, \bibinfo {author} {\bibfnamefont {G.}~\bibnamefont {Buzsaki}}, \
  and\ \bibinfo {author} {\bibfnamefont {C.}~\bibnamefont {Koch}},\ }\href@noop
  {} {\bibfield  {journal} {\bibinfo  {journal} {The Journal of Neuroscience}\
  }\textbf {\bibinfo {volume} {30}},\ \bibinfo {pages} {1925} (\bibinfo {year}
  {2010})}\BibitemShut {NoStop}%
\bibitem [{\citenamefont {Goldstein}\ \emph {et~al.}(2009)\citenamefont
  {Goldstein}, \citenamefont {Polin},\ and\ \citenamefont {Tuval}}]{Gol_09}%
  \BibitemOpen
  \bibfield  {author} {\bibinfo {author} {\bibfnamefont {R.}~\bibnamefont
  {Goldstein}}, \bibinfo {author} {\bibfnamefont {M.}~\bibnamefont {Polin}}, \
  and\ \bibinfo {author} {\bibfnamefont {I.}~\bibnamefont {Tuval}},\
  }\href@noop {} {\bibfield  {journal} {\bibinfo  {journal} {Physical Review
  Letters}\ }\textbf {\bibinfo {volume} {103}},\ \bibinfo {pages} {168103}
  (\bibinfo {year} {2009})}\BibitemShut {NoStop}%
\bibitem [{\citenamefont {Russo}\ and\ \citenamefont
  {di~Bernardo}(2009{\natexlab{a}})}]{Rus_diB_09b}%
  \BibitemOpen
  \bibfield  {author} {\bibinfo {author} {\bibfnamefont {G.}~\bibnamefont
  {Russo}}\ and\ \bibinfo {author} {\bibfnamefont {M.}~\bibnamefont
  {di~Bernardo}},\ }\href@noop {} {\bibfield  {journal} {\bibinfo  {journal}
  {Journal of Computationa Biology}\ }\textbf {\bibinfo {volume} {16}},\
  \bibinfo {pages} {379} (\bibinfo {year} {2009}{\natexlab{a}})}\BibitemShut
  {NoStop}%
\bibitem [{\citenamefont {Pecora}\ and\ \citenamefont
  {Carroll}(1990)}]{Pec_Car_90}%
  \BibitemOpen
  \bibfield  {author} {\bibinfo {author} {\bibfnamefont {L.}~\bibnamefont
  {Pecora}}\ and\ \bibinfo {author} {\bibfnamefont {T.}~\bibnamefont
  {Carroll}},\ }\href@noop {} {\bibfield  {journal} {\bibinfo  {journal}
  {Physical Review Letters}\ }\textbf {\bibinfo {volume} {64}},\ \bibinfo
  {pages} {821} (\bibinfo {year} {1990})}\BibitemShut {NoStop}%
\bibitem [{\citenamefont {Zhang}\ \emph {et~al.}(2015)\citenamefont {Zhang},
  \citenamefont {Shah}, \citenamefont {Cardenas},\ and\ \citenamefont
  {Lipson}}]{Zha_15}%
  \BibitemOpen
  \bibfield  {author} {\bibinfo {author} {\bibfnamefont {M.}~\bibnamefont
  {Zhang}}, \bibinfo {author} {\bibfnamefont {S.}~\bibnamefont {Shah}},
  \bibinfo {author} {\bibfnamefont {J.}~\bibnamefont {Cardenas}}, \ and\
  \bibinfo {author} {\bibfnamefont {M.}~\bibnamefont {Lipson}},\ }\href@noop {}
  {\bibfield  {journal} {\bibinfo  {journal} {Physical Review Letters}\
  }\textbf {\bibinfo {volume} {115}},\ \bibinfo {pages} {163902} (\bibinfo
  {year} {2015})}\BibitemShut {NoStop}%
\bibitem [{\citenamefont {Ginelli}\ \emph {et~al.}(2010)\citenamefont
  {Ginelli}, \citenamefont {Peruani}, \citenamefont {Bar},\ and\ \citenamefont
  {Chate}}]{Gin_10}%
  \BibitemOpen
  \bibfield  {author} {\bibinfo {author} {\bibfnamefont {F.}~\bibnamefont
  {Ginelli}}, \bibinfo {author} {\bibfnamefont {F.}~\bibnamefont {Peruani}},
  \bibinfo {author} {\bibfnamefont {M.}~\bibnamefont {Bar}}, \ and\ \bibinfo
  {author} {\bibfnamefont {H.}~\bibnamefont {Chate}},\ }\href@noop {}
  {\bibfield  {journal} {\bibinfo  {journal} {Physical Review Letters}\
  }\textbf {\bibinfo {volume} {104}},\ \bibinfo {pages} {184502} (\bibinfo
  {year} {2010})}\BibitemShut {NoStop}%
\bibitem [{\citenamefont {Choe}\ \emph {et~al.}(2010)\citenamefont {Choe},
  \citenamefont {Dahms}, \citenamefont {Hovel},\ and\ \citenamefont
  {Scholl}}]{Cho_10}%
  \BibitemOpen
  \bibfield  {author} {\bibinfo {author} {\bibfnamefont {C.}~\bibnamefont
  {Choe}}, \bibinfo {author} {\bibfnamefont {T.}~\bibnamefont {Dahms}},
  \bibinfo {author} {\bibfnamefont {P.}~\bibnamefont {Hovel}}, \ and\ \bibinfo
  {author} {\bibfnamefont {E.}~\bibnamefont {Scholl}},\ }\href@noop {}
  {\bibfield  {journal} {\bibinfo  {journal} {Physical Review E}\ }\textbf
  {\bibinfo {volume} {81}},\ \bibinfo {pages} {025205} (\bibinfo {year}
  {2010})}\BibitemShut {NoStop}%
\bibitem [{\citenamefont {Ng}\ and\ \citenamefont {Bassler}(2009)}]{Ng_Bas_09}%
  \BibitemOpen
  \bibfield  {author} {\bibinfo {author} {\bibfnamefont {W.}~\bibnamefont
  {Ng}}\ and\ \bibinfo {author} {\bibfnamefont {B.}~\bibnamefont {Bassler}},\
  }\href@noop {} {\bibfield  {journal} {\bibinfo  {journal} {Annual Review of
  Genetics}\ }\textbf {\bibinfo {volume} {43}},\ \bibinfo {pages} {197}
  (\bibinfo {year} {2009})}\BibitemShut {NoStop}%
\bibitem [{\citenamefont {Pesaran}\ \emph {et~al.}(2002)\citenamefont
  {Pesaran}, \citenamefont {Pezaris}, \citenamefont {Sahani}, \citenamefont
  {Mitra},\ and\ \citenamefont {Andersen}}]{Per_Pez_Sah_Mit_And_02}%
  \BibitemOpen
  \bibfield  {author} {\bibinfo {author} {\bibfnamefont {B.}~\bibnamefont
  {Pesaran}}, \bibinfo {author} {\bibfnamefont {J.}~\bibnamefont {Pezaris}},
  \bibinfo {author} {\bibfnamefont {M.}~\bibnamefont {Sahani}}, \bibinfo
  {author} {\bibfnamefont {P.}~\bibnamefont {Mitra}}, \ and\ \bibinfo {author}
  {\bibfnamefont {R.}~\bibnamefont {Andersen}},\ }\href@noop {} {\bibfield
  {journal} {\bibinfo  {journal} {Nature}\ }\textbf {\bibinfo {volume} {5}},\
  \bibinfo {pages} {805} (\bibinfo {year} {2002})}\BibitemShut {NoStop}%
\bibitem [{\citenamefont {Schlote}\ \emph {et~al.}(2013)\citenamefont
  {Schlote}, \citenamefont {Hausler}, \citenamefont {Hecker}, \citenamefont
  {Bergmann}, \citenamefont {Crisostomi}, \citenamefont {Radusch},\ and\
  \citenamefont {Shorten}}]{Sch_Sho_13}%
  \BibitemOpen
  \bibfield  {author} {\bibinfo {author} {\bibfnamefont {A.}~\bibnamefont
  {Schlote}}, \bibinfo {author} {\bibfnamefont {F.}~\bibnamefont {Hausler}},
  \bibinfo {author} {\bibfnamefont {T.}~\bibnamefont {Hecker}}, \bibinfo
  {author} {\bibfnamefont {A.}~\bibnamefont {Bergmann}}, \bibinfo {author}
  {\bibfnamefont {E.}~\bibnamefont {Crisostomi}}, \bibinfo {author}
  {\bibfnamefont {I.}~\bibnamefont {Radusch}}, \ and\ \bibinfo {author}
  {\bibfnamefont {R.}~\bibnamefont {Shorten}},\ }\href@noop {} {\bibfield
  {journal} {\bibinfo  {journal} {IEEE Transactions on Intelligent
  Transportation Systems}\ }\textbf {\bibinfo {volume} {14}},\ \bibinfo {pages}
  {1572} (\bibinfo {year} {2013})}\BibitemShut {NoStop}%
\bibitem [{\citenamefont {Russo}\ and\ \citenamefont
  {Slotine}(2010)}]{Rus_Slo_10}%
  \BibitemOpen
  \bibfield  {author} {\bibinfo {author} {\bibfnamefont {G.}~\bibnamefont
  {Russo}}\ and\ \bibinfo {author} {\bibfnamefont {J.}~\bibnamefont
  {Slotine}},\ }\href@noop {} {\bibfield  {journal} {\bibinfo  {journal}
  {Physical Review E}\ }\textbf {\bibinfo {volume} {vol. 82}},\ \bibinfo
  {pages} {041919} (\bibinfo {year} {2010})}\BibitemShut {NoStop}%
\bibitem [{\citenamefont {Bartolozzi}\ \emph {et~al.}(2005)\citenamefont
  {Bartolozzi}, \citenamefont {Leinweber},\ and\ \citenamefont
  {Thomas}}]{Bar_05}%
  \BibitemOpen
  \bibfield  {author} {\bibinfo {author} {\bibfnamefont {M.}~\bibnamefont
  {Bartolozzi}}, \bibinfo {author} {\bibfnamefont {D.~B.}\ \bibnamefont
  {Leinweber}}, \ and\ \bibinfo {author} {\bibfnamefont {A.~W.}\ \bibnamefont
  {Thomas}},\ }\href@noop {} {\bibfield  {journal} {\bibinfo  {journal}
  {Physical Review E}\ }\textbf {\bibinfo {volume} {72}},\ \bibinfo {pages}
  {046113} (\bibinfo {year} {2005})}\BibitemShut {NoStop}%
\bibitem [{\citenamefont {Wilson}\ and\ \citenamefont
  {Moehlis}(2014)}]{Wil_Moe_14}%
  \BibitemOpen
  \bibfield  {author} {\bibinfo {author} {\bibfnamefont {D.}~\bibnamefont
  {Wilson}}\ and\ \bibinfo {author} {\bibfnamefont {J.}~\bibnamefont
  {Moehlis}},\ }\href@noop {} {\bibfield  {journal} {\bibinfo  {journal}
  {Journal of Computational Neuroscience}\ } (\bibinfo {year}
  {2014})}\BibitemShut {NoStop}%
\bibitem [{\citenamefont {Wilson}\ and\ \citenamefont
  {Mo}(2014)}]{Wil_Moh_14a}%
  \BibitemOpen
  \bibfield  {author} {\bibinfo {author} {\bibfnamefont {D.}~\bibnamefont
  {Wilson}}\ and\ \bibinfo {author} {\bibfnamefont {J.}~\bibnamefont {Mo}},\
  }\href@noop {} {\bibfield  {journal} {\bibinfo  {journal} {SIAM J. Control on
  Applied Dynamical Systems}\ } (\bibinfo {year} {2014})}\BibitemShut {NoStop}%
\bibitem [{\citenamefont {Ahn}\ \emph {et~al.}(2015)\citenamefont {Ahn},
  \citenamefont {Zauber}, \citenamefont {Worth}, \citenamefont {Witt},\ and\
  \citenamefont {Rubchinsky}}]{Ahn_15}%
  \BibitemOpen
  \bibfield  {author} {\bibinfo {author} {\bibfnamefont {S.}~\bibnamefont
  {Ahn}}, \bibinfo {author} {\bibfnamefont {S.}~\bibnamefont {Zauber}},
  \bibinfo {author} {\bibfnamefont {R.}~\bibnamefont {Worth}}, \bibinfo
  {author} {\bibfnamefont {T.}~\bibnamefont {Witt}}, \ and\ \bibinfo {author}
  {\bibfnamefont {L.}~\bibnamefont {Rubchinsky}},\ }\href@noop {} {\bibfield
  {journal} {\bibinfo  {journal} {European Journal of Neuroscience}\ }
  (\bibinfo {year} {2015})}\BibitemShut {NoStop}%
\bibitem [{\citenamefont {Garcia-Ojalvo}\ \emph {et~al.}(2004)\citenamefont
  {Garcia-Ojalvo}, \citenamefont {Elowitz},\ and\ \citenamefont
  {Strogatz}}]{Gar_Elo_Str_04}%
  \BibitemOpen
  \bibfield  {author} {\bibinfo {author} {\bibfnamefont {J.}~\bibnamefont
  {Garcia-Ojalvo}}, \bibinfo {author} {\bibfnamefont {M.~B.}\ \bibnamefont
  {Elowitz}}, \ and\ \bibinfo {author} {\bibfnamefont {S.~H.}\ \bibnamefont
  {Strogatz}},\ }\href@noop {} {\bibfield  {journal} {\bibinfo  {journal}
  {Proc. of the Natl. Acad. of Sci.}\ }\textbf {\bibinfo {volume} {101}},\
  \bibinfo {pages} {10955} (\bibinfo {year} {2004})}\BibitemShut {NoStop}%
\bibitem [{\citenamefont {Tabareau}\ \emph {et~al.}(2010)\citenamefont
  {Tabareau}, \citenamefont {Slotine},\ and\ \citenamefont
  {Pham}}]{Tab_Slo_Pha_09}%
  \BibitemOpen
  \bibfield  {author} {\bibinfo {author} {\bibfnamefont {N.}~\bibnamefont
  {Tabareau}}, \bibinfo {author} {\bibfnamefont {J.}~\bibnamefont {Slotine}}, \
  and\ \bibinfo {author} {\bibfnamefont {Q.}~\bibnamefont {Pham}},\ }\href@noop
  {} {\bibfield  {journal} {\bibinfo  {journal} {PLoS Computational Biology}\
  }\textbf {\bibinfo {volume} {6}},\ \bibinfo {pages} {e1000637} (\bibinfo
  {year} {2010})}\BibitemShut {NoStop}%
\bibitem [{\citenamefont {Sakaguchi}\ and\ \citenamefont
  {Maeyama}(2013)}]{Sak_13}%
  \BibitemOpen
  \bibfield  {author} {\bibinfo {author} {\bibfnamefont {H.}~\bibnamefont
  {Sakaguchi}}\ and\ \bibinfo {author} {\bibfnamefont {S.}~\bibnamefont
  {Maeyama}},\ }\href@noop {} {\bibfield  {journal} {\bibinfo  {journal}
  {Physical Review E}\ }\textbf {\bibinfo {volume} {87}},\ \bibinfo {pages}
  {024901} (\bibinfo {year} {2013})}\BibitemShut {NoStop}%
\bibitem [{\citenamefont {Katriel}(2008)}]{Kat_08}%
  \BibitemOpen
  \bibfield  {author} {\bibinfo {author} {\bibfnamefont {G.}~\bibnamefont
  {Katriel}},\ }\href@noop {} {\bibfield  {journal} {\bibinfo  {journal}
  {Physica D}\ }\textbf {\bibinfo {volume} {237}},\ \bibinfo {pages} {2933}
  (\bibinfo {year} {2008})}\BibitemShut {NoStop}%
\bibitem [{\citenamefont {Pecora}\ and\ \citenamefont
  {Carroll}(1998)}]{Pec_Car_98}%
  \BibitemOpen
  \bibfield  {author} {\bibinfo {author} {\bibfnamefont {L.}~\bibnamefont
  {Pecora}}\ and\ \bibinfo {author} {\bibfnamefont {T.}~\bibnamefont
  {Carroll}},\ }\href@noop {} {\bibfield  {journal} {\bibinfo  {journal}
  {Physical Review E}\ }\textbf {\bibinfo {volume} {80}},\ \bibinfo {pages}
  {2019} (\bibinfo {year} {1998})}\BibitemShut {NoStop}%
\bibitem [{\citenamefont {Hu}\ \emph {et~al.}(1998)\citenamefont {Hu},
  \citenamefont {Yang},\ and\ \citenamefont {Liu}}]{Hu_Yan_Liu_98}%
  \BibitemOpen
  \bibfield  {author} {\bibinfo {author} {\bibfnamefont {G.}~\bibnamefont
  {Hu}}, \bibinfo {author} {\bibfnamefont {J.}~\bibnamefont {Yang}}, \ and\
  \bibinfo {author} {\bibfnamefont {W.}~\bibnamefont {Liu}},\ }\href@noop {}
  {\bibfield  {journal} {\bibinfo  {journal} {Physical Review E}\ }\textbf
  {\bibinfo {volume} {58}},\ \bibinfo {pages} {4440} (\bibinfo {year}
  {1998})}\BibitemShut {NoStop}%
\bibitem [{\citenamefont {l.~Huang}\ \emph {et~al.}(2009)\citenamefont
  {l.~Huang}, \citenamefont {Chen}, \citenamefont {Lai},\ and\ \citenamefont
  {Pecora}}]{Hua_Che_Lai_Pec_09}%
  \BibitemOpen
  \bibfield  {author} {\bibinfo {author} {\bibnamefont {l.~Huang}}, \bibinfo
  {author} {\bibfnamefont {Q.}~\bibnamefont {Chen}}, \bibinfo {author}
  {\bibfnamefont {Y.}~\bibnamefont {Lai}}, \ and\ \bibinfo {author}
  {\bibfnamefont {L.}~\bibnamefont {Pecora}},\ }\href@noop {} {\bibfield
  {journal} {\bibinfo  {journal} {Physical Review E}\ }\textbf {\bibinfo
  {volume} {80}},\ \bibinfo {pages} {036204} (\bibinfo {year}
  {2009})}\BibitemShut {NoStop}%
\bibitem [{\citenamefont {Pecora}\ \emph {et~al.}(2000)\citenamefont {Pecora},
  \citenamefont {Carroll}, \citenamefont {Johnson}, \citenamefont {Mar},\ and\
  \citenamefont {Fink}}]{Pec_00}%
  \BibitemOpen
  \bibfield  {author} {\bibinfo {author} {\bibfnamefont {L.}~\bibnamefont
  {Pecora}}, \bibinfo {author} {\bibfnamefont {T.}~\bibnamefont {Carroll}},
  \bibinfo {author} {\bibfnamefont {G.}~\bibnamefont {Johnson}}, \bibinfo
  {author} {\bibfnamefont {D.}~\bibnamefont {Mar}}, \ and\ \bibinfo {author}
  {\bibfnamefont {K.}~\bibnamefont {Fink}},\ }\href@noop {} {\bibfield
  {journal} {\bibinfo  {journal} {International HJournal of Bifurcations and
  Chaos}\ }\textbf {\bibinfo {volume} {10}},\ \bibinfo {pages} {273 } (\bibinfo
  {year} {2000})}\BibitemShut {NoStop}%
\bibitem [{\citenamefont {Fink}(2000)}]{Fin_00}%
  \BibitemOpen
  \bibfield  {author} {\bibinfo {author} {\bibfnamefont {K.~S.}\ \bibnamefont
  {Fink}},\ }\href@noop {} {\bibfield  {journal} {\bibinfo  {journal} {Phys.
  Rev. E, Stat. Phys. Plasmas Fluids Relat. Interdiscip. Top.}\ }\textbf
  {\bibinfo {volume} {61}},\ \bibinfo {pages} {5080 } (\bibinfo {year}
  {2000.})}\BibitemShut {NoStop}%
\bibitem [{\citenamefont {Sorrentino}(2012)}]{Sor_12}%
  \BibitemOpen
  \bibfield  {author} {\bibinfo {author} {\bibfnamefont {F.}~\bibnamefont
  {Sorrentino}},\ }\href@noop {} {\bibfield  {journal} {\bibinfo  {journal}
  {New Journal of Physics}\ }\textbf {\bibinfo {volume} {14}},\ \bibinfo
  {pages} {033035} (\bibinfo {year} {2012})}\BibitemShut {NoStop}%
\bibitem [{\citenamefont {Yang}\ \emph {et~al.}(2015)\citenamefont {Yang},
  \citenamefont {Lin}, \citenamefont {Wang},\ and\ \citenamefont
  {Huang}}]{Yan_Lin_Wan_Hua_15}%
  \BibitemOpen
  \bibfield  {author} {\bibinfo {author} {\bibfnamefont {W.}~\bibnamefont
  {Yang}}, \bibinfo {author} {\bibfnamefont {W.}~\bibnamefont {Lin}}, \bibinfo
  {author} {\bibfnamefont {X.}~\bibnamefont {Wang}}, \ and\ \bibinfo {author}
  {\bibfnamefont {L.}~\bibnamefont {Huang}},\ }\href@noop {} {\bibfield
  {journal} {\bibinfo  {journal} {Physical Review E}\ }\textbf {\bibinfo
  {volume} {91}},\ \bibinfo {pages} {032912} (\bibinfo {year}
  {2015})}\BibitemShut {NoStop}%
\bibitem [{\citenamefont {Zanette}\ and\ \citenamefont
  {Mikhailov}(1998)}]{Zan_Mik_98}%
  \BibitemOpen
  \bibfield  {author} {\bibinfo {author} {\bibfnamefont {D.}~\bibnamefont
  {Zanette}}\ and\ \bibinfo {author} {\bibfnamefont {A.}~\bibnamefont
  {Mikhailov}},\ }\href@noop {} {\bibfield  {journal} {\bibinfo  {journal}
  {Physical Review E}\ }\textbf {\bibinfo {volume} {57}},\ \bibinfo {pages}
  {276} (\bibinfo {year} {1998})}\BibitemShut {NoStop}%
\bibitem [{\citenamefont {Tass}(2001)}]{Tas_01}%
  \BibitemOpen
  \bibfield  {author} {\bibinfo {author} {\bibfnamefont {P.}~\bibnamefont
  {Tass}},\ }\href@noop {} {\bibfield  {journal} {\bibinfo  {journal}
  {Biological Cybernetics}\ }\textbf {\bibinfo {volume} {85}},\ \bibinfo
  {pages} {343} (\bibinfo {year} {2001})}\BibitemShut {NoStop}%
\bibitem [{\citenamefont {Kiss}\ \emph {et~al.}(2007)\citenamefont {Kiss},
  \citenamefont {Rusin}, \citenamefont {Kori},\ and\ \citenamefont
  {Hudson}}]{Kis_Rus_Kor_Hud_07}%
  \BibitemOpen
  \bibfield  {author} {\bibinfo {author} {\bibfnamefont {I.~Z.}\ \bibnamefont
  {Kiss}}, \bibinfo {author} {\bibfnamefont {C.~G.}\ \bibnamefont {Rusin}},
  \bibinfo {author} {\bibfnamefont {H.}~\bibnamefont {Kori}}, \ and\ \bibinfo
  {author} {\bibfnamefont {J.~L.}\ \bibnamefont {Hudson}},\ }\href@noop {}
  {\bibfield  {journal} {\bibinfo  {journal} {Science}\ }\textbf {\bibinfo
  {volume} {316}},\ \bibinfo {pages} {1886} (\bibinfo {year}
  {2007})}\BibitemShut {NoStop}%
\bibitem [{\citenamefont {Danzl}\ \emph {et~al.}(2009)\citenamefont {Danzl},
  \citenamefont {Hespanha},\ and\ \citenamefont {Moehlis}}]{Dan_Hes_Moe_09}%
  \BibitemOpen
  \bibfield  {author} {\bibinfo {author} {\bibfnamefont {P.}~\bibnamefont
  {Danzl}}, \bibinfo {author} {\bibfnamefont {J.}~\bibnamefont {Hespanha}}, \
  and\ \bibinfo {author} {\bibfnamefont {J.}~\bibnamefont {Moehlis}},\
  }\href@noop {} {\bibfield  {journal} {\bibinfo  {journal} {Biological
  Cybernetics}\ }\textbf {\bibinfo {volume} {101}},\ \bibinfo {pages} {387}
  (\bibinfo {year} {2009})}\BibitemShut {NoStop}%
\bibitem [{\citenamefont {Nabi}\ \emph {et~al.}(2013)\citenamefont {Nabi},
  \citenamefont {Mirzadeh}, \citenamefont {Gibou},\ and\ \citenamefont
  {Moehlis}}]{Nab_Mir_Gib_Moe_13}%
  \BibitemOpen
  \bibfield  {author} {\bibinfo {author} {\bibfnamefont {A.}~\bibnamefont
  {Nabi}}, \bibinfo {author} {\bibfnamefont {M.}~\bibnamefont {Mirzadeh}},
  \bibinfo {author} {\bibfnamefont {F.}~\bibnamefont {Gibou}}, \ and\ \bibinfo
  {author} {\bibfnamefont {J.}~\bibnamefont {Moehlis}},\ }\href@noop {}
  {\bibfield  {journal} {\bibinfo  {journal} {Journal of Computational
  Neuroscience}\ }\textbf {\bibinfo {volume} {34}},\ \bibinfo {pages} {259}
  (\bibinfo {year} {2013})}\BibitemShut {NoStop}%
\bibitem [{\citenamefont {Danzl}\ \emph {et~al.}(2010)\citenamefont {Danzl},
  \citenamefont {Nabi},\ and\ \citenamefont {Moehlis}}]{Dan_Nab_Moe_10}%
  \BibitemOpen
  \bibfield  {author} {\bibinfo {author} {\bibfnamefont {P.}~\bibnamefont
  {Danzl}}, \bibinfo {author} {\bibfnamefont {A.}~\bibnamefont {Nabi}}, \ and\
  \bibinfo {author} {\bibfnamefont {J.}~\bibnamefont {Moehlis}},\ }\href@noop
  {} {\bibfield  {journal} {\bibinfo  {journal} {Discrete Continuous Dynamical
  Systems}\ }\textbf {\bibinfo {volume} {28}},\ \bibinfo {pages} {1413}
  (\bibinfo {year} {2010})}\BibitemShut {NoStop}%
\bibitem [{\citenamefont {Talathi}\ \emph {et~al.}(2011)\citenamefont
  {Talathi}, \citenamefont {Carney},\ and\ \citenamefont
  {Khargonekar}}]{Tal_Car_Kha_11}%
  \BibitemOpen
  \bibfield  {author} {\bibinfo {author} {\bibfnamefont {S.}~\bibnamefont
  {Talathi}}, \bibinfo {author} {\bibfnamefont {P.}~\bibnamefont {Carney}}, \
  and\ \bibinfo {author} {\bibfnamefont {P.}~\bibnamefont {Khargonekar}},\
  }\href@noop {} {\bibfield  {journal} {\bibinfo  {journal} {Journal of
  Computational Neuroscience}\ }\textbf {\bibinfo {volume} {31}},\ \bibinfo
  {pages} {87} (\bibinfo {year} {2011})}\BibitemShut {NoStop}%
\bibitem [{\citenamefont {He}\ \emph {et~al.}(2014)\citenamefont {He},
  \citenamefont {Wang}, \citenamefont {Zhang},\ and\ \citenamefont
  {Zhan}}]{He_14}%
  \BibitemOpen
  \bibfield  {author} {\bibinfo {author} {\bibfnamefont {Z.}~\bibnamefont
  {He}}, \bibinfo {author} {\bibfnamefont {X.}~\bibnamefont {Wang}}, \bibinfo
  {author} {\bibfnamefont {G.}~\bibnamefont {Zhang}}, \ and\ \bibinfo {author}
  {\bibfnamefont {M.}~\bibnamefont {Zhan}},\ }\href@noop {} {\bibfield
  {journal} {\bibinfo  {journal} {Physical Review E}\ }\textbf {\bibinfo
  {volume} {90}},\ \bibinfo {pages} {012909} (\bibinfo {year}
  {2014})}\BibitemShut {NoStop}%
\bibitem [{\citenamefont {Heagy}\ \emph {et~al.}(1995)\citenamefont {Heagy},
  \citenamefont {Carroll},\ and\ \citenamefont {Pecora}}]{Hea_95}%
  \BibitemOpen
  \bibfield  {author} {\bibinfo {author} {\bibfnamefont {J.}~\bibnamefont
  {Heagy}}, \bibinfo {author} {\bibfnamefont {T.}~\bibnamefont {Carroll}}, \
  and\ \bibinfo {author} {\bibfnamefont {L.}~\bibnamefont {Pecora}},\
  }\href@noop {} {\bibfield  {journal} {\bibinfo  {journal} {Physical Review
  E}\ }\textbf {\bibinfo {volume} {52}},\ \bibinfo {pages} {R1253} (\bibinfo
  {year} {1995})}\BibitemShut {NoStop}%
\bibitem [{\citenamefont {de~Oliveira}\ \emph {et~al.}(2016)\citenamefont
  {de~Oliveira}, \citenamefont {Lorenzo}, \citenamefont {de~Silans},
  \citenamefont {Chevrollier}, \citenamefont {Oria},\ and\ \citenamefont
  {Cavalcante}}]{deO_16}%
  \BibitemOpen
  \bibfield  {author} {\bibinfo {author} {\bibfnamefont {G.}~\bibnamefont
  {de~Oliveira}}, \bibinfo {author} {\bibfnamefont {O.~D.}\ \bibnamefont
  {Lorenzo}}, \bibinfo {author} {\bibfnamefont {T.}~\bibnamefont {de~Silans}},
  \bibinfo {author} {\bibfnamefont {M.}~\bibnamefont {Chevrollier}}, \bibinfo
  {author} {\bibfnamefont {M.}~\bibnamefont {Oria}}, \ and\ \bibinfo {author}
  {\bibfnamefont {H.}~\bibnamefont {Cavalcante}},\ }\href@noop {} {\bibfield
  {journal} {\bibinfo  {journal} {Physical Review E}\ }\textbf {\bibinfo
  {volume} {93}},\ \bibinfo {pages} {062209} (\bibinfo {year}
  {2016})}\BibitemShut {NoStop}%
\bibitem [{\citenamefont {Vidyasagar}(1993)}]{Vid_93}%
  \BibitemOpen
  \bibfield  {author} {\bibinfo {author} {\bibfnamefont {M.}~\bibnamefont
  {Vidyasagar}},\ }\href@noop {} {\emph {\bibinfo {title} {Nonlinear systems
  analysis (2nd Ed.)}}}\ (\bibinfo  {publisher} {Pretice-Hall (Englewood
  Cliffs, NJ, USA)},\ \bibinfo {year} {1993})\BibitemShut {NoStop}%
\bibitem [{\citenamefont {Russo}\ \emph {et~al.}(2013)\citenamefont {Russo},
  \citenamefont {di~Bernardo},\ and\ \citenamefont {Sontag}}]{Rus_diB_Son_13}%
  \BibitemOpen
  \bibfield  {author} {\bibinfo {author} {\bibfnamefont {G.}~\bibnamefont
  {Russo}}, \bibinfo {author} {\bibfnamefont {M.}~\bibnamefont {di~Bernardo}},
  \ and\ \bibinfo {author} {\bibfnamefont {E.~D.}\ \bibnamefont {Sontag}},\
  }\href@noop {} {\bibfield  {journal} {\bibinfo  {journal} {IEEE Transactions
  on Automatic Control}\ }\textbf {\bibinfo {volume} {58}},\ \bibinfo {pages}
  {1328} (\bibinfo {year} {2013})}\BibitemShut {NoStop}%
\bibitem [{\citenamefont {Russo}\ and\ \citenamefont
  {Slotine}(2011)}]{Rus_Slo_11}%
  \BibitemOpen
  \bibfield  {author} {\bibinfo {author} {\bibfnamefont {G.}~\bibnamefont
  {Russo}}\ and\ \bibinfo {author} {\bibfnamefont {J.}~\bibnamefont
  {Slotine}},\ }\href@noop {} {\bibfield  {journal} {\bibinfo  {journal}
  {Physical Review E}\ }\textbf {\bibinfo {volume} {84}},\ \bibinfo {pages}
  {041929} (\bibinfo {year} {2011})}\BibitemShut {NoStop}%
\bibitem [{\citenamefont {Lohmiller}\ and\ \citenamefont
  {Slotine}(1998)}]{Loh_Slo_98}%
  \BibitemOpen
  \bibfield  {author} {\bibinfo {author} {\bibfnamefont {W.}~\bibnamefont
  {Lohmiller}}\ and\ \bibinfo {author} {\bibfnamefont {J.~J.~E.}\ \bibnamefont
  {Slotine}},\ }\href@noop {} {\bibfield  {journal} {\bibinfo  {journal}
  {Automatica}\ }\textbf {\bibinfo {volume} {34}},\ \bibinfo {pages} {683}
  (\bibinfo {year} {1998})}\BibitemShut {NoStop}%
\bibitem [{\citenamefont {Arcak}(2010)}]{Arc_10}%
  \BibitemOpen
  \bibfield  {author} {\bibinfo {author} {\bibfnamefont {M.}~\bibnamefont
  {Arcak}},\ }in\ \href@noop {} {\emph {\bibinfo {booktitle} {Proceedings of
  Amercian Control Conference}}}\ (\bibinfo {year} {2010})\BibitemShut
  {NoStop}%
\bibitem [{\citenamefont {Michel}\ \emph {et~al.}(2007)\citenamefont {Michel},
  \citenamefont {Liu},\ and\ \citenamefont {Hou.}}]{michelbook}%
  \BibitemOpen
  \bibfield  {author} {\bibinfo {author} {\bibfnamefont {A.}~\bibnamefont
  {Michel}}, \bibinfo {author} {\bibfnamefont {D.}~\bibnamefont {Liu}}, \ and\
  \bibinfo {author} {\bibfnamefont {L.}~\bibnamefont {Hou.}},\ }\href@noop {}
  {\emph {\bibinfo {title} {Stability of Dynamical Systems: Continuous,
  Discontinuous, and Discrete Systems}}}\ (\bibinfo  {publisher}
  {Springer-Verlag (New York)},\ \bibinfo {year} {2007})\BibitemShut {NoStop}%
\bibitem [{\citenamefont {Dahlquist}(1959)}]{dahlquist}%
  \BibitemOpen
  \bibfield  {author} {\bibinfo {author} {\bibfnamefont {G.}~\bibnamefont
  {Dahlquist}},\ }\href@noop {} {\emph {\bibinfo {title} {Stability and error
  bounds in the numerical integration of ordinary differential equations}}}\
  (\bibinfo  {publisher} {Transanctions of the Royal Institute Technology
  (Stockholm)},\ \bibinfo {year} {1959})\BibitemShut {NoStop}%
\bibitem [{\citenamefont {Lozinskii}(1959)}]{lozinskii}%
  \BibitemOpen
  \bibfield  {author} {\bibinfo {author} {\bibfnamefont {S.~M.}\ \bibnamefont
  {Lozinskii}},\ }\href@noop {} {\bibfield  {journal} {\bibinfo  {journal}
  {Izv. Vtssh. Uchebn. Zaved Matematika}\ }\textbf {\bibinfo {volume} {5}},\
  \bibinfo {pages} {222} (\bibinfo {year} {1959})}\BibitemShut {NoStop}%
\bibitem [{\citenamefont {Strom}(1975)}]{strom}%
  \BibitemOpen
  \bibfield  {author} {\bibinfo {author} {\bibfnamefont {T.}~\bibnamefont
  {Strom}},\ }\href@noop {} {\bibfield  {journal} {\bibinfo  {journal} {SIAM
  Journal on Numerical Analysis}\ }\textbf {\bibinfo {volume} {12}},\ \bibinfo
  {pages} {741 } (\bibinfo {year} {1975})}\BibitemShut {NoStop}%
\bibitem [{\citenamefont {Russo}\ \emph {et~al.}(2010)\citenamefont {Russo},
  \citenamefont {di~Bernardo},\ and\ \citenamefont {Sontag}}]{Rus_diB_Son_10}%
  \BibitemOpen
  \bibfield  {author} {\bibinfo {author} {\bibfnamefont {G.}~\bibnamefont
  {Russo}}, \bibinfo {author} {\bibfnamefont {M.}~\bibnamefont {di~Bernardo}},
  \ and\ \bibinfo {author} {\bibfnamefont {E.~D.}\ \bibnamefont {Sontag}},\
  }\href@noop {} {\bibfield  {journal} {\bibinfo  {journal} {PLoS Computational
  Biology}\ }\textbf {\bibinfo {volume} {6}},\ \bibinfo {pages} {e1000739}
  (\bibinfo {year} {2010})}\BibitemShut {NoStop}%
\bibitem [{\citenamefont {di~Bernardo}\ \emph {et~al.}(2014)\citenamefont
  {di~Bernardo}, \citenamefont {Liuzza},\ and\ \citenamefont
  {Russo}}]{diB_Liu_Rus_14}%
  \BibitemOpen
  \bibfield  {author} {\bibinfo {author} {\bibfnamefont {M.}~\bibnamefont
  {di~Bernardo}}, \bibinfo {author} {\bibfnamefont {D.}~\bibnamefont {Liuzza}},
  \ and\ \bibinfo {author} {\bibfnamefont {G.}~\bibnamefont {Russo}},\
  }\href@noop {} {\bibfield  {journal} {\bibinfo  {journal} {SIAM Journal on
  Control and Optimization}\ }\textbf {\bibinfo {volume} {52}},\ \bibinfo
  {pages} {3203 } (\bibinfo {year} {2014})}\BibitemShut {NoStop}%
\bibitem [{\citenamefont {Granas}(2003)}]{Gra_03}%
  \BibitemOpen
  \bibfield  {author} {\bibinfo {author} {\bibfnamefont {A.}~\bibnamefont
  {Granas}},\ }\href@noop {} {\emph {\bibinfo {title} {Fixed Point Theory}}}\
  (\bibinfo  {publisher} {Springer Verlag (New York)},\ \bibinfo {year}
  {2003})\BibitemShut {NoStop}%
\bibitem [{\citenamefont {Hartman}(1961)}]{Hartmann}%
  \BibitemOpen
  \bibfield  {author} {\bibinfo {author} {\bibfnamefont {P.}~\bibnamefont
  {Hartman}},\ }\href@noop {} {\bibfield  {journal} {\bibinfo  {journal}
  {Canadian Journal of Mathematics}\ }\textbf {\bibinfo {volume} {13}},\
  \bibinfo {pages} {480} (\bibinfo {year} {1961})}\BibitemShut {NoStop}%
\bibitem [{\citenamefont {Lewis}(1949)}]{Lewis}%
  \BibitemOpen
  \bibfield  {author} {\bibinfo {author} {\bibfnamefont {D.~C.}\ \bibnamefont
  {Lewis}},\ }\href@noop {} {\bibfield  {journal} {\bibinfo  {journal}
  {American Journal of Mathematics}\ }\textbf {\bibinfo {volume} {71}},\
  \bibinfo {pages} {294} (\bibinfo {year} {1949})}\BibitemShut {NoStop}%
\bibitem [{\citenamefont {Pavlov}\ \emph {et~al.}(2004)\citenamefont {Pavlov},
  \citenamefont {Pogromvsky}, \citenamefont {van~de Wouv},\ and\ \citenamefont
  {Nijmeijer}}]{Pav_Pog_Wou_Nij}%
  \BibitemOpen
  \bibfield  {author} {\bibinfo {author} {\bibfnamefont {A.}~\bibnamefont
  {Pavlov}}, \bibinfo {author} {\bibfnamefont {A.}~\bibnamefont {Pogromvsky}},
  \bibinfo {author} {\bibfnamefont {N.}~\bibnamefont {van~de Wouv}}, \ and\
  \bibinfo {author} {\bibfnamefont {H.}~\bibnamefont {Nijmeijer}},\ }\href@noop
  {} {\bibfield  {journal} {\bibinfo  {journal} {Systems and Control Letters}\
  }\textbf {\bibinfo {volume} {52}},\ \bibinfo {pages} {257} (\bibinfo {year}
  {2004})}\BibitemShut {NoStop}%
\bibitem [{\citenamefont {Angeli}(2002)}]{Ang_02}%
  \BibitemOpen
  \bibfield  {author} {\bibinfo {author} {\bibfnamefont {D.}~\bibnamefont
  {Angeli}},\ }\href@noop {} {\bibfield  {journal} {\bibinfo  {journal} {IEEE
  Transactions on Automatic Control}\ }\textbf {\bibinfo {volume} {47}},\
  \bibinfo {pages} {410} (\bibinfo {year} {2002})}\BibitemShut {NoStop}%
\bibitem [{\citenamefont {Lohmiller}\ and\ \citenamefont
  {Slotine}(2005)}]{pde}%
  \BibitemOpen
  \bibfield  {author} {\bibinfo {author} {\bibfnamefont {W.}~\bibnamefont
  {Lohmiller}}\ and\ \bibinfo {author} {\bibfnamefont {J.~J.~E.}\ \bibnamefont
  {Slotine}},\ }\href@noop {} {\bibfield  {journal} {\bibinfo  {journal}
  {International Journal of Control}\ }\textbf {\bibinfo {volume} {78}},\
  \bibinfo {pages} {678} (\bibinfo {year} {2005})}\BibitemShut {NoStop}%
\bibitem [{\citenamefont {Jouffroy}(2005)}]{jou_05}%
  \BibitemOpen
  \bibfield  {author} {\bibinfo {author} {\bibfnamefont {J.}~\bibnamefont
  {Jouffroy}},\ }in\ \href@noop {} {\emph {\bibinfo {booktitle} {Proceedings of
  the International Conference on Decision and Control}}}\ (\bibinfo {year}
  {2005})\ pp.\ \bibinfo {pages} {5450-- 5455}\BibitemShut {NoStop}%
\bibitem [{\citenamefont {Aminzare}\ and\ \citenamefont
  {Sontag}(2014{\natexlab{a}})}]{Ami_Son_14}%
  \BibitemOpen
  \bibfield  {author} {\bibinfo {author} {\bibfnamefont {Z.}~\bibnamefont
  {Aminzare}}\ and\ \bibinfo {author} {\bibfnamefont {E.}~\bibnamefont
  {Sontag}},\ }\href@noop {} {\bibfield  {journal} {\bibinfo  {journal} {IEEE
  Transactions on Network Science and Engineering}\ }\textbf {\bibinfo {volume}
  {91}},\ \bibinfo {pages} {91 } (\bibinfo {year}
  {2014}{\natexlab{a}})}\BibitemShut {NoStop}%
\bibitem [{\citenamefont {Aminzare}\ and\ \citenamefont
  {Sontag}(2014{\natexlab{b}})}]{Ami_Son_14a}%
  \BibitemOpen
  \bibfield  {author} {\bibinfo {author} {\bibfnamefont {Z.}~\bibnamefont
  {Aminzare}}\ and\ \bibinfo {author} {\bibfnamefont {E.}~\bibnamefont
  {Sontag}},\ }in\ \href@noop {} {\emph {\bibinfo {booktitle} {Proceedings of
  the IEEE Conference on Decision and Control}}}\ (\bibinfo {year}
  {2014})\BibitemShut {NoStop}%
\bibitem [{\citenamefont {Horn}\ and\ \citenamefont
  {Johnson}(1999)}]{Hor_Joh_99}%
  \BibitemOpen
  \bibfield  {author} {\bibinfo {author} {\bibfnamefont {R.~A.}\ \bibnamefont
  {Horn}}\ and\ \bibinfo {author} {\bibfnamefont {C.~R.}\ \bibnamefont
  {Johnson}},\ }\href@noop {} {\emph {\bibinfo {title} {Matrix Analysis}}}\
  (\bibinfo  {publisher} {Cambridge University Press (Cambridge, UK)},\
  \bibinfo {year} {1999})\BibitemShut {NoStop}%
\bibitem [{\citenamefont {Watts}\ and\ \citenamefont
  {Strogatz}(1998)}]{Wat_Str_98}%
  \BibitemOpen
  \bibfield  {author} {\bibinfo {author} {\bibfnamefont {D.}~\bibnamefont
  {Watts}}\ and\ \bibinfo {author} {\bibfnamefont {S.}~\bibnamefont
  {Strogatz}},\ }\href@noop {} {\bibfield  {journal} {\bibinfo  {journal}
  {Nature}\ }\textbf {\bibinfo {volume} {393}},\ \bibinfo {pages} {440}
  (\bibinfo {year} {1998})}\BibitemShut {NoStop}%
\bibitem [{\citenamefont {Strogatz}(2003)}]{Str_03}%
  \BibitemOpen
  \bibfield  {author} {\bibinfo {author} {\bibfnamefont {S.}~\bibnamefont
  {Strogatz}},\ }\href@noop {} {\emph {\bibinfo {title} {Sync: the emerging
  science of spontaneous order}}}\ (\bibinfo  {publisher} {Hyperion (New York,
  USA)},\ \bibinfo {year} {2003})\BibitemShut {NoStop}%
\bibitem [{\citenamefont {Yagamuchi}\ \emph {et~al.}(2003)\citenamefont
  {Yagamuchi}, \citenamefont {Isejima}, \citenamefont {Matsuo}, \citenamefont
  {Okura},\ and\ \citenamefont {Yagita}}]{Yag_Ise_Mat_Oku_Yag_03}%
  \BibitemOpen
  \bibfield  {author} {\bibinfo {author} {\bibfnamefont {S.}~\bibnamefont
  {Yagamuchi}}, \bibinfo {author} {\bibfnamefont {H.}~\bibnamefont {Isejima}},
  \bibinfo {author} {\bibfnamefont {T.}~\bibnamefont {Matsuo}}, \bibinfo
  {author} {\bibfnamefont {R.}~\bibnamefont {Okura}}, \ and\ \bibinfo {author}
  {\bibfnamefont {K.}~\bibnamefont {Yagita}},\ }\href@noop {} {\bibfield
  {journal} {\bibinfo  {journal} {Science}\ }\textbf {\bibinfo {volume}
  {302}},\ \bibinfo {pages} {2531 } (\bibinfo {year} {2003})}\BibitemShut
  {NoStop}%
\bibitem [{\citenamefont {Russo}\ and\ \citenamefont
  {di~Bernardo}(2009{\natexlab{b}})}]{Rus_diB_09}%
  \BibitemOpen
  \bibfield  {author} {\bibinfo {author} {\bibfnamefont {G.}~\bibnamefont
  {Russo}}\ and\ \bibinfo {author} {\bibfnamefont {M.}~\bibnamefont
  {di~Bernardo}},\ }\href@noop {} {\bibfield  {journal} {\bibinfo  {journal}
  {IEEE Transactions on Circuit and Systems II}\ }\textbf {\bibinfo {volume}
  {56}},\ \bibinfo {pages} {177} (\bibinfo {year}
  {2009}{\natexlab{b}})}\BibitemShut {NoStop}%
\bibitem [{\citenamefont {Ripperger}\ and\ \citenamefont
  {Albrecht}(2012)}]{Rip_12}%
  \BibitemOpen
  \bibfield  {author} {\bibinfo {author} {\bibfnamefont {J.}~\bibnamefont
  {Ripperger}}\ and\ \bibinfo {author} {\bibfnamefont {U.}~\bibnamefont
  {Albrecht}},\ }\href@noop {} {\bibfield  {journal} {\bibinfo  {journal} {Cell
  Research}\ } (\bibinfo {year} {2012})}\BibitemShut {NoStop}%
\bibitem [{\citenamefont {Szallasi}\ \emph {et~al.}(2006)\citenamefont
  {Szallasi}, \citenamefont {Stelling},\ and\ \citenamefont
  {Periwal}}]{Sza_Ste_Per_06}%
  \BibitemOpen
  \bibfield  {author} {\bibinfo {author} {\bibfnamefont {Z.}~\bibnamefont
  {Szallasi}}, \bibinfo {author} {\bibfnamefont {J.}~\bibnamefont {Stelling}},
  \ and\ \bibinfo {author} {\bibfnamefont {V.}~\bibnamefont {Periwal}},\
  }\href@noop {} {\emph {\bibinfo {title} {System Modeling in Cellular Biology:
  From Concepts to Nuts and Bolts}}}\ (\bibinfo  {publisher} {The MIT Press},\
  \bibinfo {year} {2006})\BibitemShut {NoStop}%
\bibitem [{\citenamefont {Del~Vecchio}\ \emph {et~al.}(2008)\citenamefont
  {Del~Vecchio}, \citenamefont {Ninfa},\ and\ \citenamefont
  {Sontag}}]{DelV_Nin_Son_08}%
  \BibitemOpen
  \bibfield  {author} {\bibinfo {author} {\bibfnamefont {D.}~\bibnamefont
  {Del~Vecchio}}, \bibinfo {author} {\bibfnamefont {A.~J.}\ \bibnamefont
  {Ninfa}}, \ and\ \bibinfo {author} {\bibfnamefont {E.~D.}\ \bibnamefont
  {Sontag}},\ }\href@noop {} {\bibfield  {journal} {\bibinfo  {journal} {Nature
  Molecular Systems Biology}\ }\textbf {\bibinfo {volume} {4}},\ \bibinfo
  {pages} {161} (\bibinfo {year} {2008})}\BibitemShut {NoStop}%
\bibitem [{\citenamefont {Arnold}(1978)}]{Arn_78}%
  \BibitemOpen
  \bibfield  {author} {\bibinfo {author} {\bibfnamefont {V.~I.}\ \bibnamefont
  {Arnold}},\ }\href@noop {} {\emph {\bibinfo {title} {Mathematical methods of
  classical mechanics}}}\ (\bibinfo  {publisher} {Spriger-Verlag (New York)},\
  \bibinfo {year} {1978})\BibitemShut {NoStop}%
\bibitem [{\citenamefont {Wang}\ and\ \citenamefont
  {Slotine}(2005)}]{Wan_Slo_05}%
  \BibitemOpen
  \bibfield  {author} {\bibinfo {author} {\bibfnamefont {W.}~\bibnamefont
  {Wang}}\ and\ \bibinfo {author} {\bibfnamefont {J.~J.~E.}\ \bibnamefont
  {Slotine}},\ }\href@noop {} {\bibfield  {journal} {\bibinfo  {journal}
  {Biological Cybernetics}\ }\textbf {\bibinfo {volume} {92}},\ \bibinfo
  {pages} {38} (\bibinfo {year} {2005})}\BibitemShut {NoStop}%
\end{thebibliography}
\end{document}